\documentclass[12pt,preprint]{aastex}
\usepackage{natbib}
\usepackage{url}
\usepackage{amsmath}
\usepackage{appendix}
\usepackage{hyperref}

\begin{document}

\title{Highly Embedded $8\mu$m cores of Star Formation in the Spiral Arms and
Filaments of 15 Nearby Disk Galaxies}

\author{Bruce G. Elmegreen\altaffilmark{1}, Debra Meloy Elmegreen\altaffilmark{2}}


\altaffiltext{1}{IBM Research Division, T.J. Watson Research Center, 1101 Kitchawan
Road, Yorktown Heights, NY 10598; bge@us.ibm.com}

\altaffiltext{2}{Department of Physics \& Astronomy, Vassar College, Poughkeepsie,
NY 12604}

\begin{abstract}
Spitzer Space Telescope observations of 15 spiral galaxies show numerous dense
cores at $8\mu$m that are revealed primarily in unsharp mask images. The
cores are generally invisible in optical bands because of extinction, and they are
also indistinct at $8\mu$m alone because of contamination by more widespread
diffuse emission. Several hundred core positions, magnitudes, and colors from the
four IRAC bands are measured and tabulated for each galaxy. The larger galaxies,
which tend to have longer and more regular spiral arms, often have their infrared
cores aligned along these arms, with additional cores in spiral arm spurs.
Galaxies without regular spirals have their cores in more irregular spiral-like
filaments, with typically only one or two cores per filament. Nearly every
elongated emission feature has $8\mu$m cores strung out along its length. The
occurrence of dense cores in long and thin filaments is reminiscent of filamentary
star formation in the solar neighborhood, although on a scale 100 times larger in
galaxies. The cores most likely form by gravitational instabilities and cloud
agglomeration in the filaments. The simultaneous occurrence of several cores with
regular spacings in some spiral arms suggests that in these cases, all of the
cores formed at about the same time and the corresponding filaments are young.
Total star formation rates for the galaxies correlate with the total embedded
stellar masses in the cores with an average ratio corresponding to a possible age
between 0.2 Myr and 2 Myr. This suggests that the identified cores are the
earliest phase for most star formation.
\end{abstract}

\section{Introduction}
\label{intro}

Spitzer Space Telescope InfraRed Array Camera (IRAC) images of M100 show regularly
spaced $8\mu$m peaks in the main spiral arms, suggesting that star formation began
by gravitational collapse in dense gas compressed into thin dust lanes by stellar
spiral arm shocks \citep[][hereafter Paper I]{elmegreen18}. Such shocks were
originally modeled by \cite{roberts69}. Their gravitational fragmentation into
massive cloud complexes was considered early on by \cite{elmegreen79},
\cite{cowie81}, \cite{balbus85}, and \cite{tomisaka87}. Cloud-forming instabilities
inside swing-amplified spirals, which are themselves the result of separate
instabilities in two dimensions \citep{goldreich65,toomre81}, were shown
analytically for magnetic gas by \cite{elmegreen94} and simulated by \cite{kim01},
\cite{kim02a} and \cite{kim07}. Spiral arm formation was recently reviewed by
\cite{dobbs14} and \cite{shu16}.

A related process is the formation of interarm spurs downstream from an unstable
spiral arm shock \citep{balbus88}, and the fragmentation of these spurs to form
massive clouds, as modeled by \cite{kim02b}. In three-dimensional models of local
density wave shocks by \cite{kim06}, spur formation begins inside the shock with
regular condensations and then evolves to make more condensations in the spurs
downstream. Galaxy-wide spur formation in two dimensions was first simulated by
\cite{chakrabarti03}, who emphasized ultraharmonic resonances, and \cite{shetty06},
who highlighted the magneto-hydrodynamic instabilities.  Regularly-spaced spiral arm
condensations and spurs were also found by \cite{dobbs08} in both their
non-self-gravitating and self-gravitating models; at low gas surface density
($4\;M_\odot$ pc$^{-2}$), the condensations formed by cloud agglomeration with a
characteristic separation given by the epicycle radius, and at high surface density
($20\;M_\odot$ pc$^{-2}$) they formed by gravitational instabilities that enhanced
and focussed cloud agglomeration. \citep{khoperskov13} modeled cloud formation and
molecular chemistry in self-gravitating gas shocked by stellar spiral arms; cloud
formation was initiated by shear instabilities and then cloud growth followed by
agglomeration.

Self-gravitating simulations at sub-pc resolution by \cite{renaud13,renaud14} made
spiral arms with regularly spaced gas clouds; they suggested that the arms
fragmented by gravitational instabilities, although possible contributions from the
Kelvin-Helmholtz instability were considered also \citep[related to the wiggle
instability in][]{wada04}. Analytical models of magnetic spiral arm instabilities
are in \cite{lee12} and \cite{inoue18,inoue19}. Simulations of spiral arms in
proto-planetary disks also show collapse into cores \citep{kratter16}.

Gravitational instabilities in local filaments
\citep[e.g.,][]{rossano78,schneider79,andre17}, including long Milky Way dust clouds
that may be spiral arm shocks \citep[e.g.][]{goodman14, ragan14, mattern18}, should
have a fastest growing mode consisting of regularly-spaced condensations with a
characteristic separation proportional to the filament width, depending on magnetic
field strength and the equation of state \citep[e.g.,][]{chandra53, stod63,
nagasawa87,inutsuka92,nakamura93,tomisaka95,fiege00,seo16,
kainulainen16,clarke17,lou17, hoss18,chira18}. This regularity in spiral arms was
first measured in 23 galaxies by \cite{elmegreen83}, who concluded that the
associated cloud complexes have masses comparable to the turbulent Jeans mass of
$\sim10^7\;M_\odot$. Such giant clouds were also identified in the first quadrant of
the Milky Way by \cite{elmegreen87}, following earlier work on regularly spaced
giant HI clouds in the outer Milky Way by \cite{mcgee64}. Recent studies by
\cite{koo18} traced these clouds in the Perseus and Outer Milky Way arms.

Most giant molecular clouds in the Milky Way are clustered together in the inner
regions of giant HI clouds that are also distributed along the spiral arms
\citep{elmegreen87,grabelsky87}, suggesting there is a cascade of collapse from
spiral arm shocks to dense star-forming cores. Observations of GMCs alone or of the
interstellar medium on scales of $\sim1$ kpc or less do not usually reveal the
larger regularities that might be associated with spiral arms, nor do conventional
observations of star formation in other galaxies, where extinction and young stellar
feedback can confuse the process. Other studies of this regularity include the giant
star-forming regions in a long and thin tidal shock at mid-radius in the interacting
galaxies NGC 2207 and IC 2163 \citep{elmegreen06}, and in the spiral arms of M31
\citep{efremov10} and NGC 628 \citep{gusev13}, which is also one of our galaxies
here.

This paper examines 15 galaxies to look for small infrared emission peaks like those
in M100. They are revealed here primarily in unsharp mask images made by dividing
the IRAC $8\mu$m image at $2.4^{\prime\prime}$ (FWHM) resolution by a blurred
version of the same image. This division removes the extended emission. Section
\ref{method} discusses our method for identification and measurement, Section
\ref{properties} compiles observations of the core colors and magnitudes, Section
\ref{comp} compares core properties for different types of galaxies, Section
\ref{implications} discusses the implications for star formation, and Section
\ref{conc} summarizes our conclusions.

\section{Observations}
\label{method}

\begin{figure}
\epsscale{1.}
\includegraphics[width=5.5in]{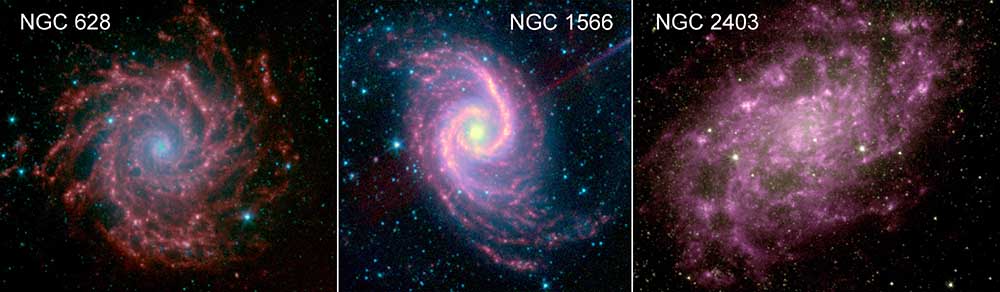}\\
\includegraphics[width=5.5in]{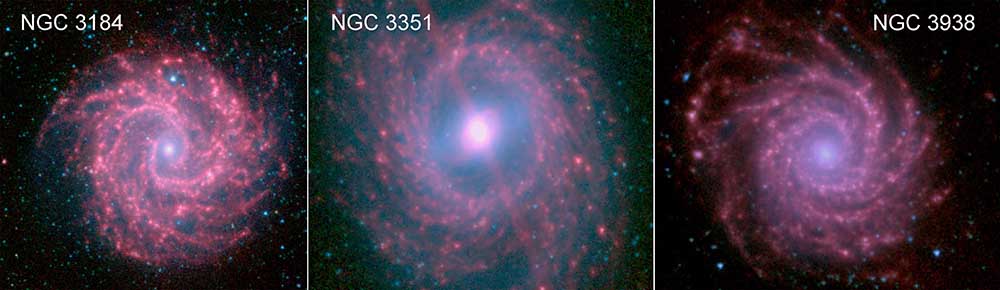}\\
\includegraphics[width=5.5in]{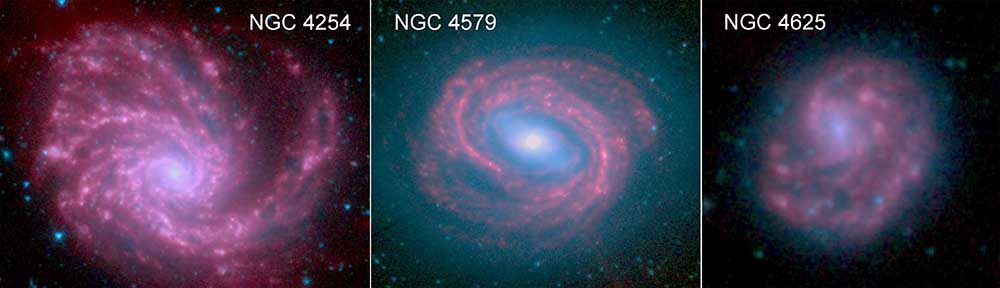}\\
\includegraphics[width=5.5in]{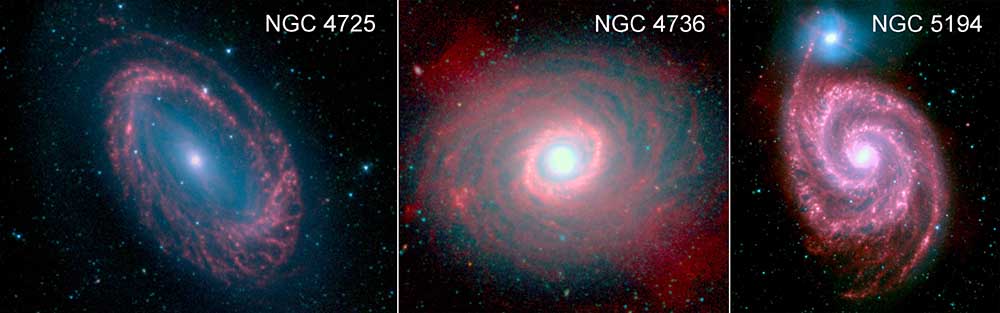}\\
\includegraphics[width=5.5in]{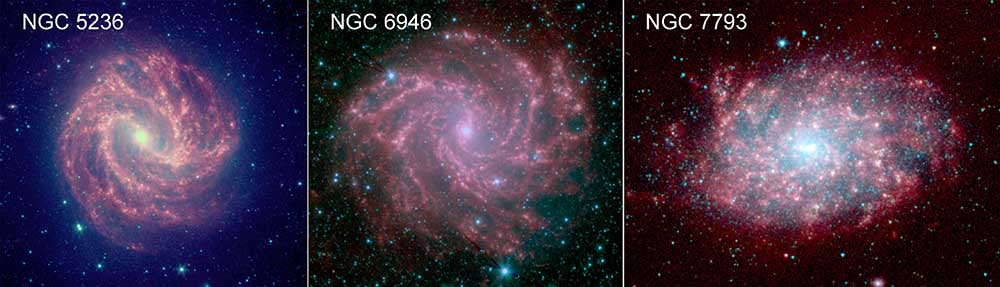}\\
\caption{Color IRAC images of the 15 galaxies in this study from composites
of $3.6\mu$m (in blue), $4.5\mu$m (in green), and $8\mu$m (in red) images. [Figure quality
reduced for arXiv.]}
\label{mycolor_all}
\end{figure}

Galaxies in the SINGS \citep{kennicutt03} and KINGFISH \citep{kennicutt11} surveys
plus M83 (NGC 5236) from the Spitzer Space Telescope archives were selected for low
inclination, large angular size, and diversity of spiral structure. The galaxies
span Hubble types from Sab through Sm. Spiral arm classes, which generally correlate
with galaxy mass, span the range from flocculent to multiple arm to grand design.
The post-Basic Calibrated Data (BCD) processed images in all four IRAC bands
($3.6\mu$m, $4.5\mu$m, $5.8\mu$m, and $8.0\mu$m) and the $24\mu$m MIPS images were
retrieved as FITS files and then aligned through the Image Reduction Analysis
Facility (IRAF) task {\it wregister}. Figure \ref{mycolor_all} shows the 15 galaxies
with color IRAC images made using the software package DS9. Each is a composite of
the archived $3.6\mu$m (in blue), $4.5\mu$m (in green),  and $8\mu$m (in red)
images. Table \ref{galaxylist} lists key properties of the 15 chosen galaxies, with
GSR distance, distance modulus, Hubble type, and radius $R_{25}$ in arcmin and kpc
from the NASA/IPAC Extragalactic Database
(NED)\footnote{https://ned.ipac.caltech.edu/}. The table also includes the parsec
scale for one arcsecond, Arm Class from \cite{elmegreen87b}, star formation rate
(SFR) from \cite{thilker07} or \cite{kennicutt11} as indicated, and number of chosen
cores as discussed below. The scale indicates that even though the features found
here are relatively small in our images, corresponding to $3^{\prime\prime}$
diameters, they are still large compared to giant molecular clouds in most galaxies.

\begin{figure}
\epsscale{1.}
\includegraphics[width=6.in]{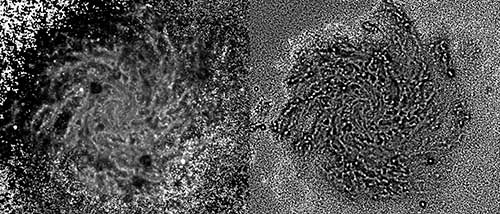}
\caption{Unsharp masks made by dividing the $8\mu$m image of NGC 628 by the $24\mu$m image (left)
and by dividing the same $8\mu$m image by a Gaussian-blurred version of the $8\mu$m image (right).
The version on the right has
better definition of the clumps, so we use that technique here. [Figure quality
reduced for arXiv.]}
\label{N628ch4divmips24_and_N628ch4divg3}
\end{figure}

In a first step of our analysis, the $8\mu$m images were divided by the
corresponding $24\mu$m images using the IRAF task {\it imarith}, and displayed as
grayscale images in DS9. This was the procedure used in Paper I. This process
results in a type of unsharp mask image, since the lower resolution of the
$24\mu$m image compared with the $8\mu$m is approximately equivalent to blurring the
original $8\mu$m image. A small adjustment of contrast and brightness then usually
reveals a network of small bright cores in the arms, spurs, shells, and other
filamentary structures. The result for NGC 628 is on the left in Figure
\ref{N628ch4divmips24_and_N628ch4divg3}.  Some of the $8\mu$m cores have large
black regions around them where the $24\mu$m emission was particularly bright or
large compared to the $8\mu$m emission. Recall that the $24\mu$m images have an
angular resolution three larger than the $8\mu$m images, so features with the same
physical size and brightness in the two frames would end up looking like a top hat,
i.e., a peak in the center the size of the $8\mu$m resolution ($2.4^{\prime\prime}$)
and a valley in a ring around the center the size of the $24\mu$m resolution
($7.1^{\prime\prime}$).

A more revealing procedure is an unsharp mask made by dividing the $8\mu$m
image by a blurred version of the same $8\mu$m image using the IRAF task ${\it
gauss}$ with a 3 pixel {\bf sigma}. The right-hand panel in Figure
\ref{N628ch4divmips24_and_N628ch4divg3} shows the result of this.  Clearly the
small-scale $8\mu$m structure shows up better than in the $8\mu$m/$24\mu$m image
because the background is more uniform. The cores are present in both but in the
outer regions they are more evident in the right-hand image. In what follows, we use
this second procedure to identify $8\mu$m cores.

Unsharp masks are a highly selective process, designed to emphasize a certain
angular scale. Nevertheless, we use this process because the features of interest
are not easily seen in the $8\mu$m or $24\mu$m images alone (nor in the other IRAC
bands), and they are also not in optical images (Paper I). This does not mean they
cannot be seen at $8\mu$m; in fact most individual cores can be seen as peaks in the
$8\mu$ image if the contrast is tuned for that core, but this optimum contrast is
generally different for each $8\mu$m emission region because the peak intensities
and background levels differ. Thus, division by the $8\mu$m or $24\mu$m images
also normalizes the backgrounds to an approximately constant value, and then the IR
cores are all visible in a single image where their relative positions and
importance for star formation can be assessed.

For each galaxy, we measured clearly visible $8\mu$m cores in all four IRAC
passbands. The selection effects in unsharp masking imply that distribution
functions for luminosity, mass, and size only show the selected parts of more
complete functions. For example, the selected core luminosities and sizes increase
with galaxy distance (Sect. \ref{comp}). Still, the tabulated positions of the
$8\mu$m cores, and their relationships to spiral arms and other structures, can
provide clues about their formation. These tabulations can also be a useful guide to
other observations, such as molecular gas at arcsecond resolution, which will be
necessary to study the associated cloud densities and kinematics.

In total for the 15 galaxies, over 8300 IR cores were identified. The IRAF
task {\it phot} was used with DS9 to find the position of each $8\mu$m peak,
manually moving the computer cursor to each one and placing it in the central pixel
of the peak as viewed in the magnified cutout of the peak's neighborhood.  A list of
all the selected positions was made in this way, followed by a plot of these
positions as blue dots on a blank field. This plot was then overlayed on the unsharp
mask image using Adobe Photoshop, after scaling it to the same size so that each
plotted dot covered the $8\mu$m peak. The overlay was examined for peaks that were
overlooked, and the positions of these peaks were measured and added to the list.
After a few iterations of this procedure, most of the small $8\mu$m peaks were noted
and tabulated. We avoided choosing extended or diffuse emission regions but only
picked the small, essentially unresolved peaks. Automatic searching for the peaks
with SExtractor failed to find them consistently, as they all have different
backgrounds and intensities and are often parts of larger features and filaments
that SExtractor finds instead.

Figures \ref{compositegalaxies1} to \ref{compositegalaxies5} show unsharp mask
images with and without blue dots, using selection criteria for $8\mu$m cores that
are discussed in Section \ref{properties}. A comparison between the two images
reveals a tiny $8\mu$m emission peak in the left-hand figure for each blue dot in the
right-hand figure. Emission peaks without blue dots were either too faint to be
systematically studied or star-like (see below).

\section{Properties of $8\mu$m Cores}
\label{properties}

Magnitudes of the $8\mu$m peaks were determined in the four IRAC bands using the
IRAF task {\it phot} with a measurement aperture of 2 pixels radius
($3.0^{\prime\prime}$ diameter, which encompasses the IR cores that stand out in the
unsharp mask images), and background subtraction from an annulus between 3 and 4
pixels away from the core center. The zeropoints for conversion of counts to
Vega-system magnitudes in each filter were taken from the IRAC Instrument
handbook\footnote{http://irsa.ipac.caltech.edu/data/SPITZER/docs/irac/irac
instrumenthandbook/}. The right ascension, declination, magnitude and error for each
filter, along with the colors for each core, are listed in Tables
\ref{listofclumps628} to \ref{listofclumps7793}.

The distributions of $8\mu$m apparent magnitudes are shown in Figure
\ref{efremov_mags}. They are in a fairly narrow range, from about 13 to 15 mag, for
all of the galaxies as a result of the unsharp mask selection.  The distributions
have long tails at the faint ends that are unphysical and suffer from incompleteness
and confusion with noise. To have a representative catalog for future use, we placed
a lower limit on the peak brightness of 16 mag, which is 1 to 3 magnitudes fainter
than the peak. The measured cores that are fainter than these cutoffs (and other
cores that are discarded as well, see below) have a cyan color for that part of the
histogram in the figure.  The red histograms for NGC 1566 and NGC 5194 are for
the cores in the two main spiral arms, which are brighter than the others (Sect.
\ref{sparm}).

Measurements of NGC 4625 using a 3 pixel radius aperture with a background
annulus of 4 to 5 pixels, instead of an aperture radius of 2 pixels as in the
figures, had slightly brighter magnitudes (less than in proportion to the larger
areas) because most of the cores are extended with diffuse infra-red emission
surrounding them. Thus the tabulated magnitudes do not reflect the total light from
the whole star-forming region but only the light from the inner $4\pi$ square pixels
of area, which corresponds to 7.07 arcsec$^2$. They are therefore related to the
core central surface brightness, which equals $[8.0]+2.12$ in units of mag
arcsec$^{-2}$, where $[8.0]$ is the core apparent magnitude at $8\mu$m.

Color-color plots in Figure \ref{efremov_colors} are not subject to the selection
effects of unsharp masking (nor to distance effects) but nonetheless show a narrow
range. The average values for the Vega-system colors and their dispersions $\sigma$
for all the galaxies are $[3.6]-[4.5]=0.21\pm0.007$, $\sigma([3.6]-[4.5])=0.53$,
$[4.5]-[5.8]=2.15\pm0.01$, $\sigma([4.5]-[5.8])=0.79$, and
$[5.8]-[8.0]=1.81\pm0.005$, $\sigma([5.8]-[8.0])=0.39$. These colors are the same as
what we found for compact IRAC sources in the interacting galaxies NGC 2207 and IC
2163 \citep{elmegreen06}, and by unsharp masking using $8\mu$m and $24\mu$m images
for M100 \citep{elmegreen18}.

The IRAC colors suggest that emission from extincted photospheres makes
$[3.6]-[4.5]$ positive but low in the Vega system, combined with emission from PAHs
to make $[5.8]-[8.0]$ high. Bare photospheric emission has a negative slope in the
$3.6\mu$m to $4.5\mu$m range \citep{li01} with $[3.6]-[4.5]\sim-0.5$ in the AB
system \cite[e.g.,][]{stern05} and $[3.6]-[4.5]\sim0$ in the Vega system
\citep{allen04}. Extinction is needed to bring the average color up to our measured
Vega mag value of $[3.6]-[4.5]=0.21\pm0.008$. Considering that a visual extinction
of 30 mag produces a $[3.6]-[4.5]$ color excess between 0.4 and 0.45 depending on
the extinction law \citep{megeath04}, the average visual extinction for our sources
is about half of this, or $\sim15$ mag. This extinction corresponds to a near-side
gas mass surface density of $\Sigma_{\rm gas}\sim300\;M_\odot$ pc$^{-2}$, using the
local conversion factor between color excess and H I column density from
\cite{bohlin78}, $E(B-V)=N(HI)/(5.8\times10^{21}\; {\rm cm}^{-2})$, and a ratio of
total-to-selective extinction $A_{\rm V}/E(B-V)=3.1$ with a mean molecular weight of
1.36 times the hydrogen mass.  If the gas mass is the same on the far side of the
extincted source as on the near side, then the total gas mass surface density could
be $\sim600\;M_\odot$ pc$^{-2}$. If an embedded source with this extinction and
surface density occupies a fraction $f$ of the $3^{\prime\prime}$ diameter of our
photometry measurements, and the typical conversion to length is pc/arcsec$\sim50$
from Table \ref{galaxylist}, then the corresponding gas mass is
$\sim10^7f^2\;M_\odot$.

Extinction barely affects the $[5.8]-[8.0]$ color \citep{allen04}, so the IRAC
emission from our sources also requires a significant excess at $8\mu$m. This could
not from be Class II protostars, for example, which have extincted photospheres and
about the right $[3.6]-[4.5]$, because such protostars have disks that are too warm,
making $[5.8]-[8.0]$ too blue: \cite{allen04} and \cite{megeath04} show that Vega
$[3.6]-[4.5]\sim0-0.8$ and $[5.8]-[8.0]\sim0.4-1.1$ for this class. They could be
Class 0 protostars, which have extincted photospheres with cool envelopes and disks;
\citet[][see their Fig. 7]{whitney03} model low-inclination members of this class
with $[3.6]-[4.5]\sim0.5-0.9$ and $[5.8]-[8.0]\sim1.2-1.5$. They also note that
planetary nebulae and reflection nebulae have about the same colors. However, large
dusty regions like we observe are not likely to have all their IRAC bands dominated
by any of these sources, not the Class 0 because these stars are usually short-lived
and rare compared to other classes, and not planetary or reflection nebulae.

A clue to the high $[5.8]-[8.0]$ color of our cores comes from whole galaxies, which
have about the same value from PAH emission that peaks at $\sim8\mu$m
\citep{stern05,winston07,gutermuth09,stutz13}. Galaxies have lower $[3.6]-[4.5]$
than the cores we observe because most of their photospheric emission is not highly
extincted. The most likely explanation for the observed core colors, considering the
presence of our $8\mu$m peaks in dust filaments, is that the low $[3.6]-[4.5]$ comes
from extincted photospheres of an embedded cluster, and the high $[5.8]-[8.0]$ comes
from strong PAH emission on carbonaceous grains, which produce an emission bump at
$6\mu$m to $10\mu$m \citep{li01}.

Some galaxies like NGC 6946 also had a dozen or more peaks at $([3.6]-[4.5],
[5.8]-[8.0])\sim(0,0) $ in Vega mag. These objects are likely stellar \citep
{allen04}, so we placed a lower cutoff at $[5.8]-[8.0]=0.6$ to avoid them. Any core
that had no measurable emission in one or more of the four IRAC bands was discarded
as well.

In summary, the right-hand sides of Figures \ref{compositegalaxies1} to
\ref{compositegalaxies5}, and Tables \ref{listofclumps628} to
\ref{listofclumps7793}, contain all of the small peaks that we examined on the
unsharp mask images which have apparent $8\mu$m magnitudes brighter than 16,
measurable emission in each IRAC band, and a $[5.8]-[8.0]$ color that exceeds $0.6$.
This represents a total of 6315 objects.

\section{Comparisons among different galaxies and spiral arm types}
\label{comp}

\begin{figure}
\epsscale{1.}
\includegraphics[width=5.5in]{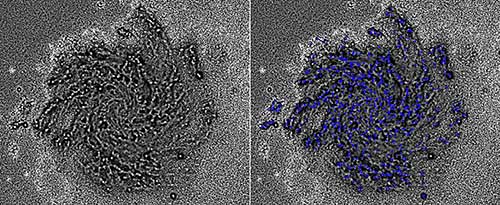}\\
\includegraphics[width=5.5in]{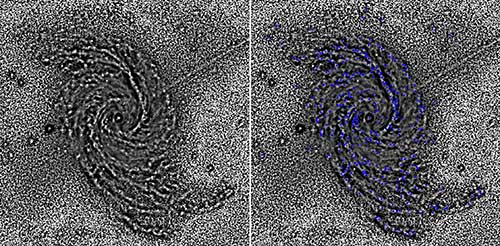}\\
\includegraphics[width=5.5in]{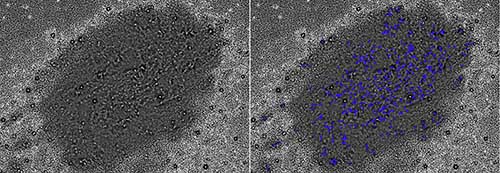}\\
\caption{Unsharp mask images of NGC 628 (top), NGC 1566 (middle), and NGC 2403 (bottom).
The figures on the right show
blue dots for the chosen cores that are also listed in the tables. These cores have
an apparent $8\mu$m magnitude brighter than $16$ mag and a $[3.4]-[4.5]$ color greater than
0.6 to avoid unextincted stars. Star-like objects with no blue dots have been excluded because
of one of these three criteria. [Figure quality
reduced for arXiv.]}
\label{compositegalaxies1}
\end{figure}

\begin{figure}
\epsscale{1.}
\includegraphics[width=5.2in]{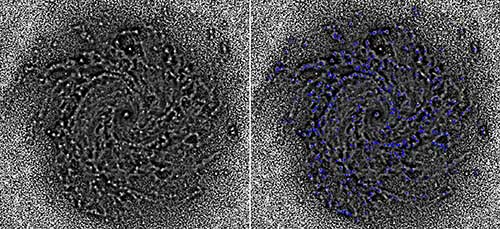}\\
\includegraphics[width=5.2in]{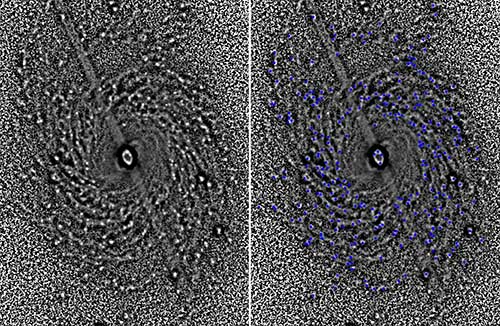}\\
\includegraphics[width=5.2in]{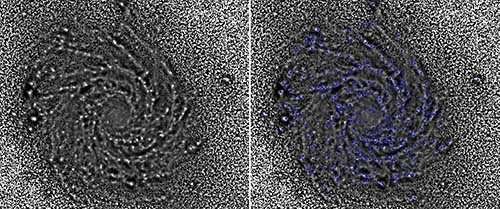}\\
\caption{Unsharp mask images of NGC 3184 (top), NGC 3351 (middle), and NGC 3938 (bottom). [Figure quality
reduced for arXiv.]}
\label{compositegalaxies2}
\end{figure}

\begin{figure}
\epsscale{1.}
\includegraphics[width=5.5in]{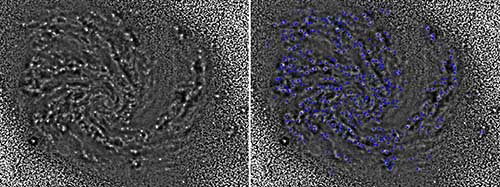}\\
\includegraphics[width=5.5in]{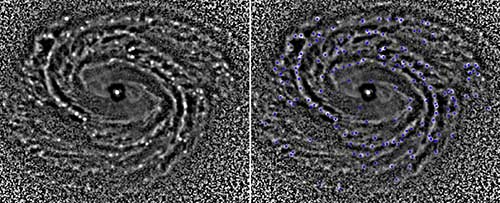}\\
\includegraphics[width=5.5in]{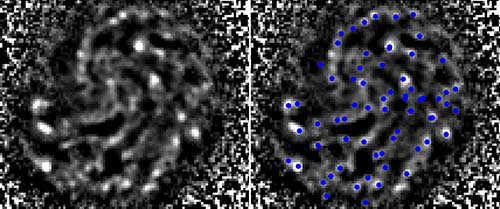}\\
\caption{Unsharp mask images of NGC 4254 (top), NGC 4579 (middle), and NGC 4625 (bottom). [Figure quality
reduced for arXiv.]}
\label{compositegalaxies3}
\end{figure}

\begin{figure}
\epsscale{1.}
\includegraphics[width=5.in]{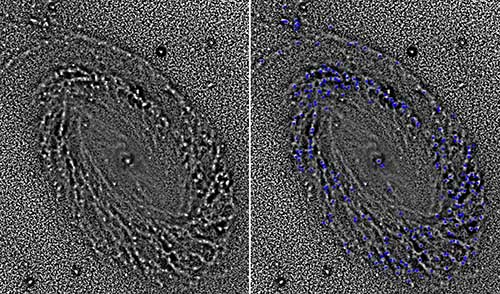}\\
\includegraphics[width=5.in]{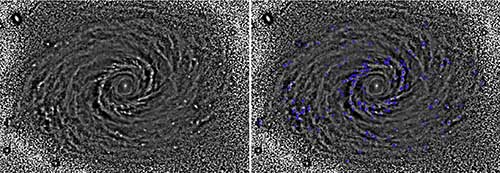}\\
\includegraphics[width=5.in]{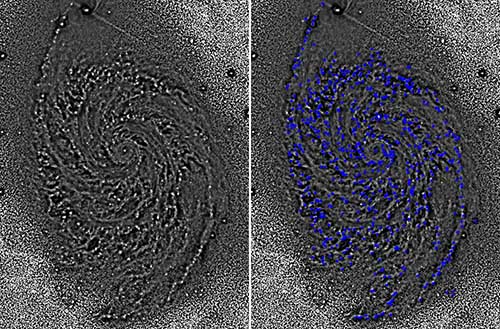}\\
\caption{Unsharp mask images of NGC 4725 (top), NGC 4736 (middle), and NGC 5194 (bottom). [Figure quality
reduced for arXiv.]}
\label{compositegalaxies4}
\end{figure}

\begin{figure}
\epsscale{1.}
\includegraphics[width=5.5in]{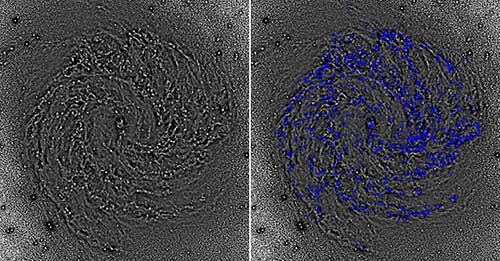}\\
\includegraphics[width=5.5in]{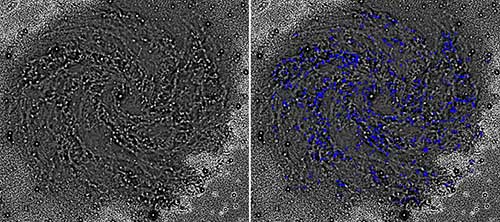}\\
\includegraphics[width=5.5in]{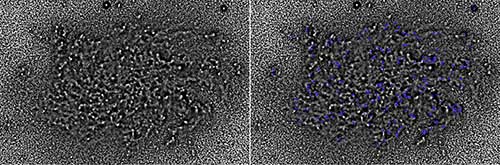}\\
\caption{Unsharp mask images of NGC 5236 (top), NGC 6946 (middle), and NGC 7793 (bottom). [Figure quality
reduced for arXiv.]}
\label{compositegalaxies5}
\end{figure}

\begin{figure}
\epsscale{1.}
\includegraphics[width=6.in]{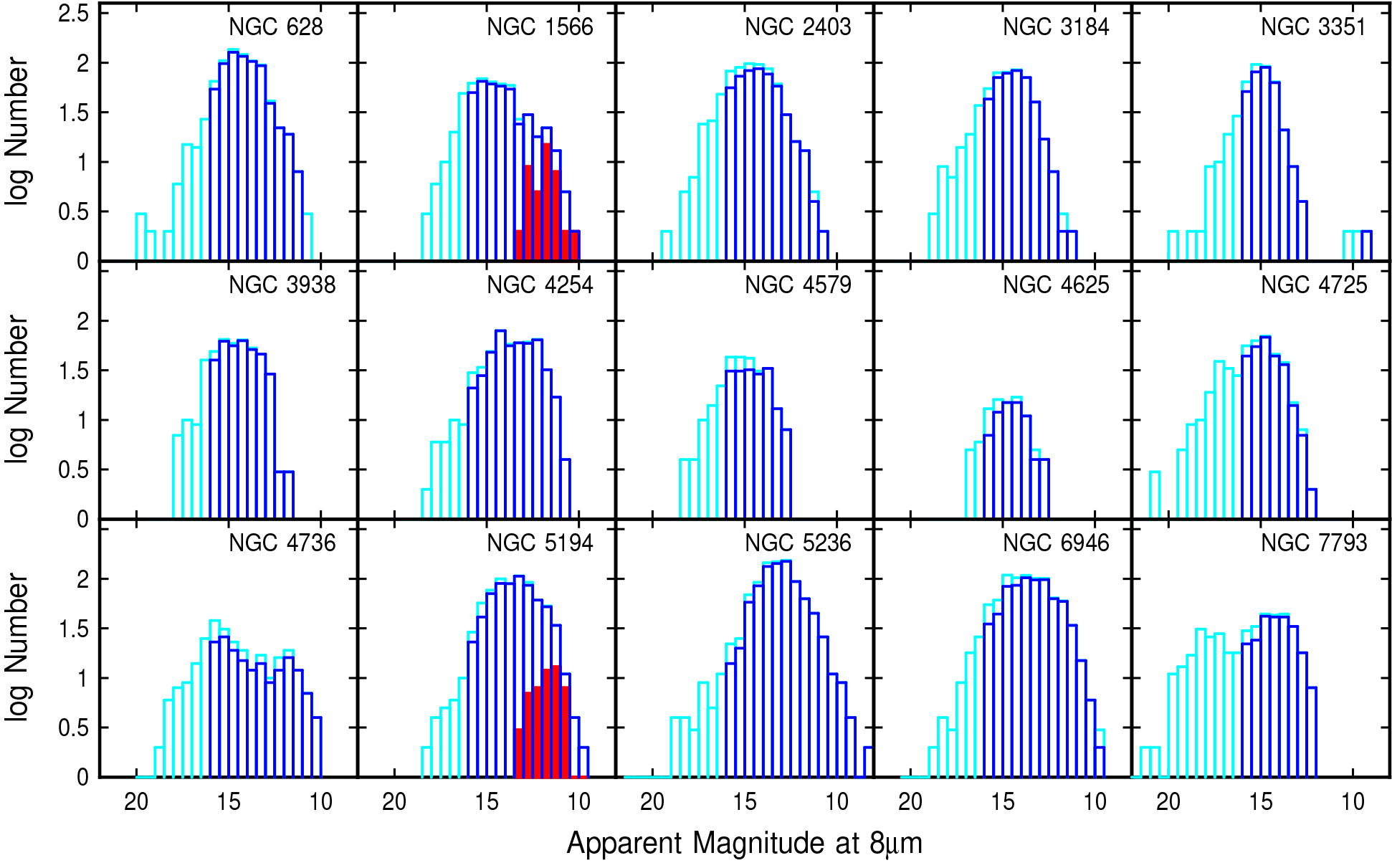}
\caption{Distribution of apparent magnitudes at $8\mu$m for all measured cores. Blue
histograms are for cores that satisfy our three cutoff limits: brighter than 16 mag,
detection in all IRAC passbands, and non-stellar ($[5.8]-[8.0]>0.6$).
Cyan histograms are for measured cores that fail one or more of these conditions and are
included here for completeness. Red histograms are for clumps in the main spiral arms of
the two galaxies with grand design spiral structure. }
\label{efremov_mags}
\end{figure}

\begin{figure}
\epsscale{1.}
\includegraphics[width=6.in]{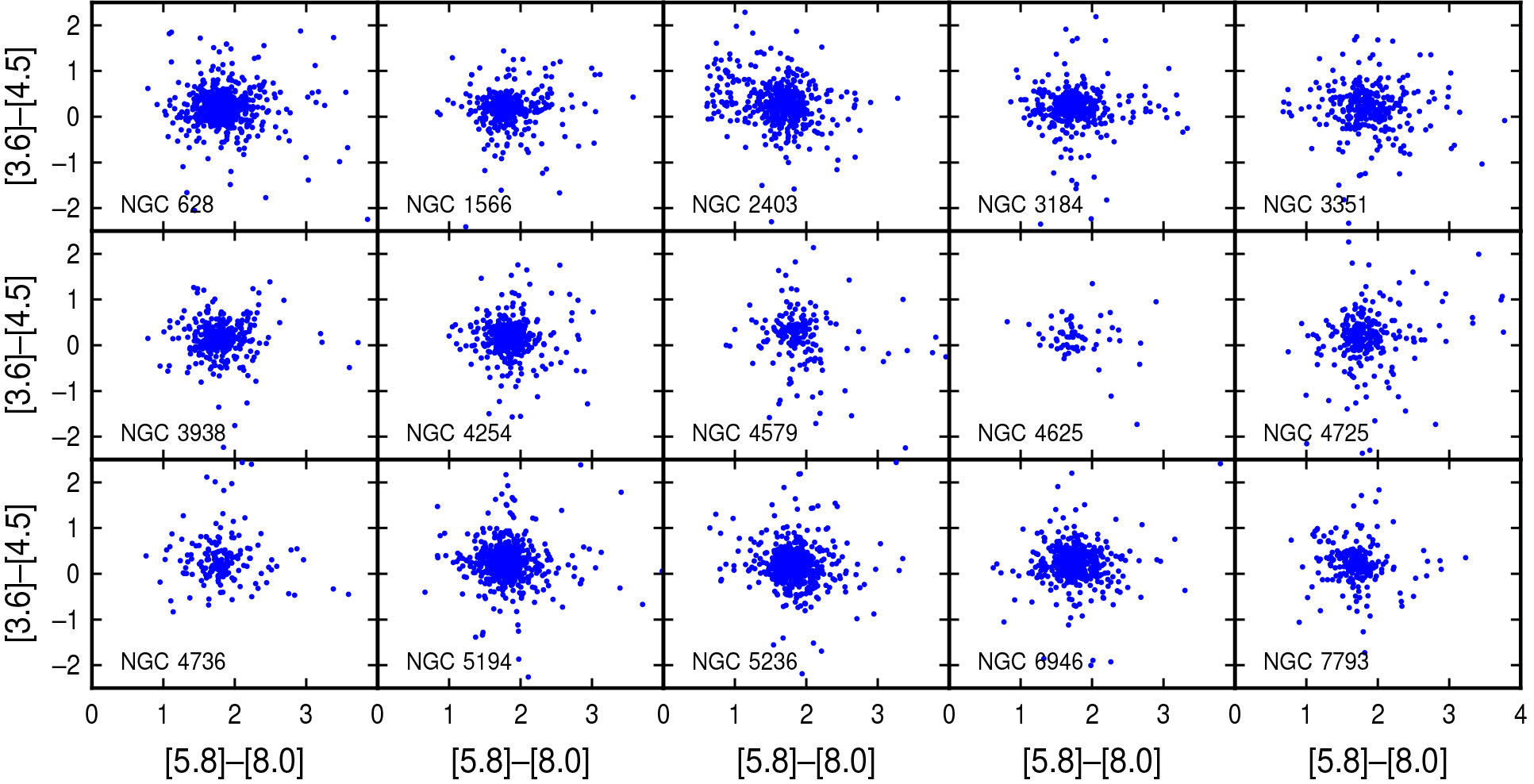}
\caption{Color-color diagrams of selected cores. The colors are about the same for
each galaxy and suggest embedded clusters with $\sim15$ mag of visual extinction and
strong PAH emission.}
\label{efremov_colors}
\end{figure}

Comparison of the color IRAC images in Figure \ref{mycolor_all} and the unsharp mask
images in Figures \ref{compositegalaxies1}-\ref{compositegalaxies5} shows how the
unsharp mask images diminish the relative brightness of the main spiral arms and
emphasize the bright and faint features equally. NGC 1566 and NGC 5194 are good
examples, as the unsharp mask images barely show the well-known two-arm spirals that
dominate the optical and near-infrared images.

There are other, more subtle, differences too, such as the small white dots that are
in the unsharp mask images but indistinct in the IRAC color images. These are
$8\mu$m cores that are probably compact, highly-extincted regions of recent star
formation (Sect. \ref{properties}). Most of the thin spiral arms in all our galaxies
have strings of cores all along their lengths. These cores are sometimes regularly
spaced but not always. The spiral arm spurs generally have a few cores too, as do
occasional rings in the outer regions, which might be supershells.

The cores are typically separated by 5 to 20 pixels ($3.75^{\prime\prime}$ to
$15^{\prime\prime}$), with more cores in longer filaments. Intrinsic core sizes are
not known because the $\sim2.4^{\prime\prime}$ FWHM of the $8\mu$m image typically
corresponds to $\sim150$ pc in these galaxies (see the scale in Table
\ref{galaxylist}), and most embedded star-forming cores in resolved studies are
smaller than this.  Diffuse patches of emission in the unsharp mask images generally
do not have cores, so a threshold surface density is apparently required to make
them. The correspondence between the $8\mu$m cores and the spiral arms and other
features is discussed briefly here in order of NGC number for the galaxies.

As shown in Figure \ref{compositegalaxies1}, NGC 628 is a multiple arm galaxy with
strings of irregularly-spaced cores lying along all of the many arms and spurs and a
few isolated cores between them in short filaments. NGC 1566 is a grand design
galaxy in optical images and a multiple-arm galaxy in the unsharp mask image. It has
cores along all of the thin arms and in several short spurs, but relatively few
cores in the thick patches of diffuse emission. NGC 2403 is a flocculent, low-mass
galaxy with short spiral arms and a few irregularly spaced cores in each of them.

In Figure \ref{compositegalaxies2}, NGC 3184 is a multiple arm galaxy with two long
strings of cores in the inner arms and a few cores in spurs and irregular filaments
between the arms. NGC 3351 is an inner ring galaxy with many arms. Few regular
strings of $8\mu$m cores appear, but the cores that are there are sprinkled
throughout the arms and spurs. NGC 3938 is a multiple arm galaxy whose arms are
filled with strings of cores. Very few show up between the main arms, except in
spurs.

In Figure \ref{compositegalaxies3}, NGC 4254 has multiple thick arms irregularly
filled with cores. There are ring-like structures in the outer regions that also
contain cores. NGC 4579 is a multiple arm galaxy with narrow arms sparsely filled
with cores; there are relatively few spurs and most do not contain cores. NGC 4625
is a late-type multiple arm galaxy dominated by short arms, most of which have a few
cores with relatively large separations.

In Figure \ref{compositegalaxies4}, NGC 4725 has an inner ring and flocculent
structure like 3351. There are long strings of cores along some of the arms and many
short arms with just a few cores. NGC 4736 is an outer ring flocculent galaxy; its
narrow arms are feathery and contain very few cores compared with all of the other
galaxies in our sample. NGC 5194, a grand design galaxy dominated by two arms in the
optical, appears as a multiple arm galaxy in this image. The two dominant arms are
not obvious, and numerous cores are in both the arms and the spurs.

In Figure \ref{compositegalaxies5}, NGC 5236 (with a streak coming from the $24\mu$m
image) is a multiple arm galaxy with many cores in the arms and very few between the
arms. NGC 6946 is a multiple arm galaxy with bright arms in the color image but more
irregular structure in the unsharp mask. There are many cores, both in the arms and
spurs with mostly irregular spacings. NGC 7793, another flocculent low-mass galaxy,
shows irregularly placed cores.

NGC 7793, NGC 2403, and NGC 6946 are all at about the same distance, where the
photometric aperture used for the IR core measurements ($3^{\prime\prime}$) covers
the same physical diameter, $\sim50$ pc. The different numbers and
distributions of $8\mu$m cores in these galaxies are therefore not an artifact of
distance. The most distant galaxy in our sample, NGC 4254, still shows $8\mu$m cores
lining the spiral arms.

The cores in the main arms of the two galaxies with the brightest two-arm
spirals, NGC 1566 and NGC 5194, are brighter than the cores outside the arms. The
brightness distributions of these spiral arm cores are shown in Figure
\ref{efremov_mags} as red bars (these cores are also included in the total counts at
these magnitudes, shown by the blue lines in the histogram).  The average $8\mu$m
magnitude of 53 main spiral arm cores in NGC 1566 is 11.92 and the average outside
the main arms is 14.04. For 43 main arm cores in NGC 5194, these averages are 11.70
and 13.32. The magnitude of the average of the core fluxes (i.e., $-2.5\log_{10}$ of
the average of $10^{-0.4[8.0]}$ for magnitudes [8.0]) is 11.68 in the main arms of
NGC 1566 and 12.94 outside the main arms. For NGC 5194, these are 11.41 and 12.62.
Considering these latter averages, the ratio of core brightness in the arms to
outside the arms is 3.2 and 3.0, respectively. The same ratio was found in M100
(Paper I). Star-forming regions in spiral arms are well known to be brighter and
more massive than elsewhere \citep{rozas96}.

The similar appearance of the cores, nearly independent of distance, is partly a
selection effect because all of the galaxies have about the same angular size and
therefore pixel count in their diameters.  The average and dispersion of the galaxy
diameter distribution in pixels are 696 px and 281 px, respectively.  Thus bigger
galaxies have larger distances between the cores, making the images look similar.
This scaling of galaxy morphology is a well-known phenomenon
\citep[e.g.,][]{block84} and is not a bias in our selection of cores at the limiting
magnitude. Bigger, more distant galaxies would presumably have more cores at fainter
magnitudes than we can observe.

\begin{figure}
\epsscale{1.}
\includegraphics[width=4.in]{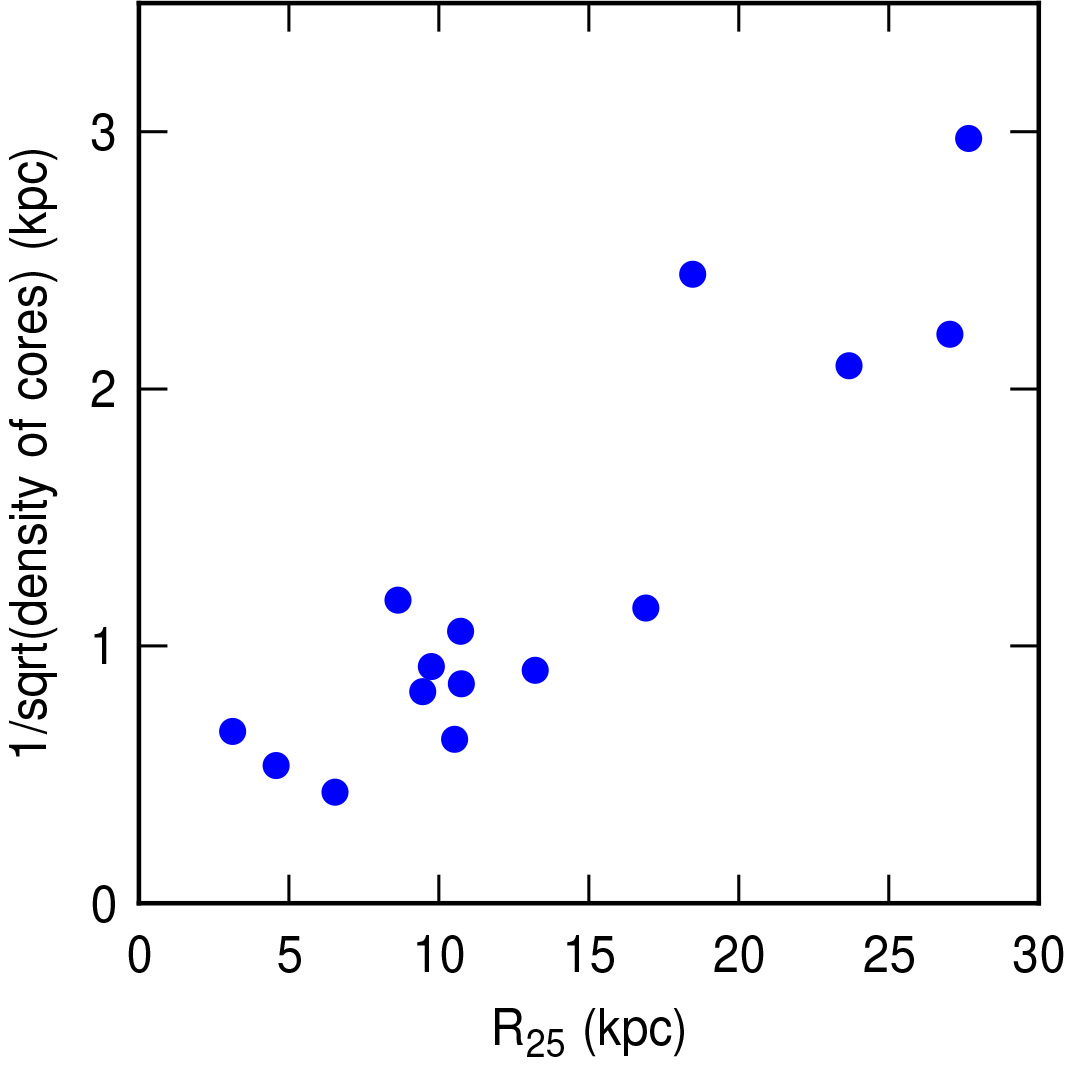}
\caption{The mean physical separation between chosen cores scales with the galaxy
size, indicating a self-similar nature to star forming regions given our selection effects,
as explained by equation 2.}
\label{efremov_table1}
\end{figure}

Figure \ref{efremov_table1} plots the inverse square root of the average density of
cores in a galaxy, calculated as the number of cores divided by the area out to
$R_{25}$, versus $R_{25}$. This abscissa is the average separation between cores.
The linear relation in the figure shows that the average separation scales with the
galaxy size, which is the same as saying that the number of chosen cores is about
constant, as also evident from Table \ref{galaxylist}.

The approximately constant number of cores exceeding an apparent magnitude limit is
consistent with a common luminosity function for equal-age star clusters, which is
the same as the mass function. Suppose the luminosity function is
$dn(L)/dL=n_0L^{-2}$. If we define a maximum luminosity $L_{\rm max}$ such that the
integral over the luminosity function from $L_{\rm max}$ to infinity is unity, then
$n_0=L_{\rm max}$. The total number of cores exceeding some minimum luminosity is
\begin{equation}
N(L>L_{\rm min})=L_{\rm max}\int_{L_{\rm min}}^\infty L^{-2}dL =L_{\rm max}/L_{\rm min}
\end{equation}
for $L_{\rm max}>>L_{\rm min}$. At constant core density and constant core spacing,
and with a constant minimum luminosity for core definition, this total number should
increase with galaxy area, $\pi R_{25}^2$ for $R_{25}$ in kpc. Now consider a
limiting luminosity for observation, $L_{\rm obs}$. For a fixed apparent magnitude,
this limiting absolute luminosity increases as the square of the distance to the
galaxy, $D$. The number of cores expected to be seen is thus
\begin{equation}
N(L>L_{\rm obs})=L_{\rm max}\int_{L_{\rm obs}}^\infty L^{-2}dL =L_{\rm max}/L_{\rm obs}.
\end{equation}
The numerator in this expression increases as $R_{25}^2$ and the denominator
increases as $D^2$. For an approximately constant galaxy angular size, as is
typically chosen for surveys, $R_{25}\propto D$.  Thus, the number of cores above a
fixed apparent magnitude is independent of distance, and their average spacing
scales with the galaxy size.  This result implies that if one of our distant
galaxies could be observed closer, it would have a larger angular size and more
cores above our fixed apparent magnitude limit. Also, the intrinsically fainter
cores (not visible in our survey with the true galaxy distance) would probably be
located between the cores we observe now in the distant galaxy, with a shorter
spacing between then, measured in kpc, for the nearby version.

Also because of the limited angular resolution, the cores could be subdivided into
multiple molecular clouds and star-forming regions. The lack of specific optical
emission from many of these cores suggests that even if they are fragmented, they
are still unlikely to have a high fraction of their young stars in the
low-extinction regions between the fragments.

\section{Implications for Star Formation}
\label{implications}
\subsection{Spiral Arms}
\label{sparm}

The core colors in IRAC (Fig. \ref{efremov_colors}) suggest highly embedded star
formation with significant PAH emission (Sect. \ref{properties}). Because many are
located in dust ridges on the inner parts of spiral arms, they are probably evidence
for star formation initiated by a density wave shock, as suggested in general terms
by \cite{roberts69}. There are many early models for how such triggering might
occur, including cloud collisions \citep{kwan83,scoville83} and the Parker
instability \citep{mouschovias74}, but the occasional appearance of regularity and
the relative importance of gravity in the Parker-Jeans instability
\citep{elmegreen82,chou00} suggests they formed by some mixture of agglomeration and
magneto-gravitational instabilities in the dense gas. Presumably the embedded stars
will break out of their clouds, estimated above to have a mean visible extinction of
$\sim15$ mag on the near side of the galaxy disk, and then they will become visible
as HII regions, OB associations and young optical clusters. Subsequently, they will
age as they move downstream from the shock, causing a blue-to-red color gradient, as
found in some cases \citep[e.g.,][]{martinez09,shabani18,peterken19,yu19}.

This standard scenario is not always expected. Some spiral arms are not
quasi-stationary stellar waves, as envisioned by Lin and Shu, but transient spirals
growing from two-dimensional disk instabilities and wrapping up in a shear flow
\citep[e.g.,][]{baba13}. Then the gas and young stars do not move much relative to
the stellar spiral as they age, and color gradients should be small or even reversed
\citep{dobbs10,shabani18}.

Because we find $8\mu$m cores in spiral arm spurs and at shell edges downstream from
the main dust lanes (Sect. \ref{comp}), star formation continues even as the gas
moves through the arm. Obviously, this decreases the average age gradient too,
producing young stars all throughout the spiral and even in the trailing interarm
region.  The $8\mu$ cores in spurs and shells also suggest a likely role for
self-gravity in forming these objects.

Most $8\mu$m cores are not regularly spaced along their dusty filaments, although
some are over short segments (see Figs.
\ref{compositegalaxies1}-\ref{compositegalaxies5}).  Regions with regular spacing
suggest an instability or forced-agglomeration of smaller clouds, with a spacing at
the fastest growing mode along the filament (Sect. \ref{intro}). They also suggest
that the whole segment with the regularity started this process at about the same
time. Otherwise, the filaments should be more irregular with the youngest regions
forming wherever the gas is just now building up and the older regions empty where a
previous generation just broke out. This lack of simultaneity could be the reason
most filaments do not have regularly spaced cores.

Transient spirals also prolong the time when the gas is dense and can form
condensations by instabilities. In the model by \cite{dobbs08a}, the gas collected
in the arms and stayed there while the spiral grew and moved. There was no age
gradient and no spurs because there was no downstream flow of the gas through the
arms.

\subsection{A possible connection to the Jeans mass}

\begin{figure}
\epsscale{1.3}
\includegraphics[width=6.in]{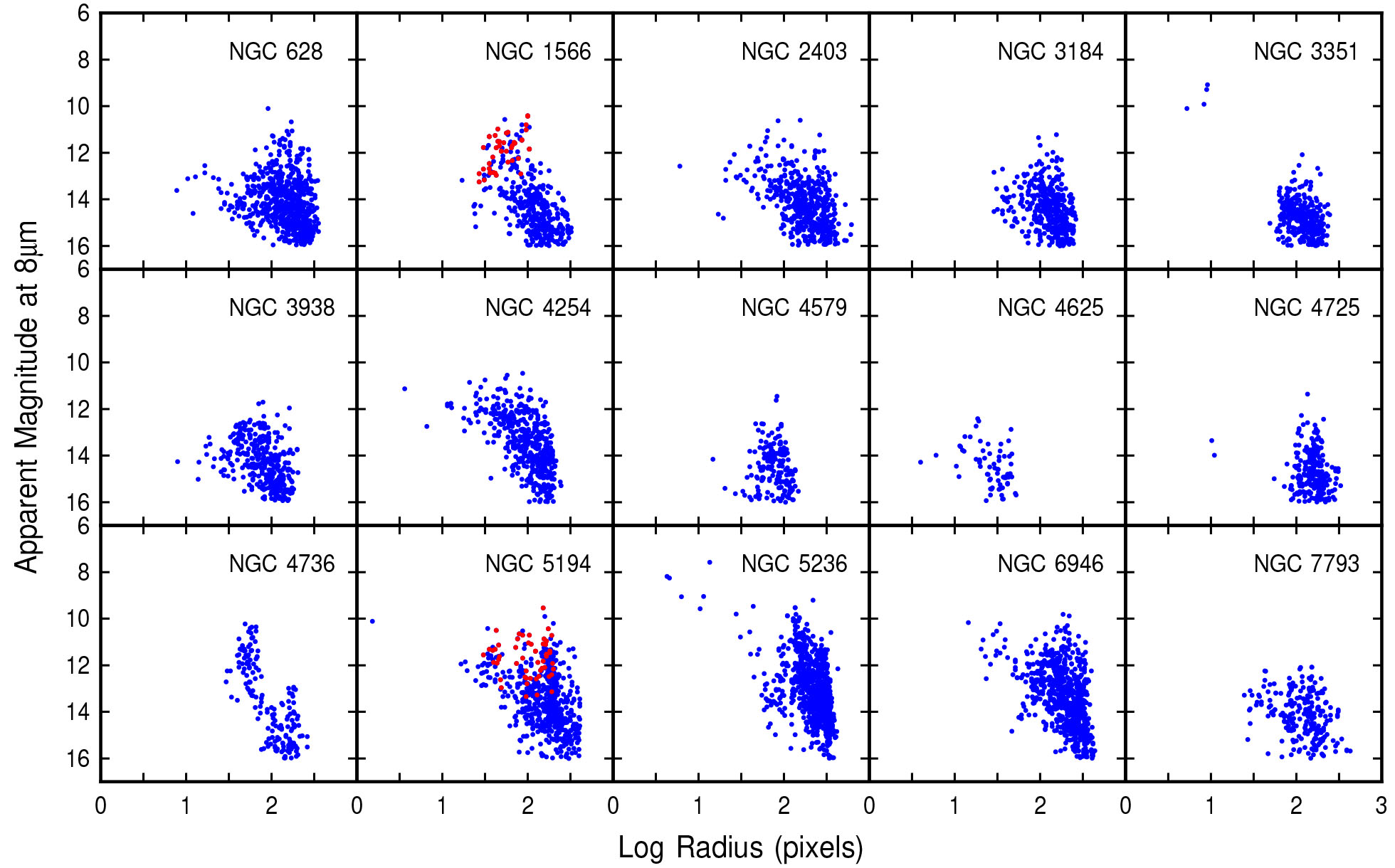}
\caption{Clump magnitude plotted as a function of galactocentric radius, in pixels. The red points
are in the spiral arms of NGC 1566 and NGC 5194. The lower
left boundaries of the distributions could be a selection effect where faint clumps are not
clearly visible in the inner region where the disk is brighter,
or they could be an indication of a decreasing Jeans mass with radius, as expected
for a relatively constant disk thickness. The increasing brightness with radius for spiral arm clumps
in NGC 1566 could indicate an increasing disk thickness in the arms.}
\label{efremov_mag_radial}
\end{figure}

To determine whether the $8\mu$m core luminosities might be related to a
gravitational Jeans mass, we plotted the core magnitudes versus the distances from
the galaxy's centers. The approximately exponential decrease in gas surface density
with radius, $\Sigma_{\rm gas}(R)$, and the relatively constant gas thickness before
the outer flare region \citep[$H(R)$,][]{wilson19}, suggests the Jeans mass, which
is approximately $\Sigma_{\rm gas}H^2$, should decrease with galactocentric radius.
Figure \ref{efremov_mag_radial} shows the results for our cores. Red points are for
spiral arm cores, as also distinguished by red histograms in Figure
\ref{efremov_mags}. There is a large range of core magnitudes for most galaxies in
the outer regions, but only bright cores are in the inner regions, considering the
radial trend in the lower envelop of points. This trend is consistent with the
expected trend in Jeans mass, but it may also arise from a selection effect where
faint cores in the inner regions do not stand out significantly above the brighter
disks there and so are not selected in our survey. The unsharp mask technique makes
the average core backgrounds uniform, but not the noise in those backgrounds, which
scales approximately with the square root of the pixel count. The actual trend in
the figure looks more like the effect of a Jean's mass, however. Typically the
faintest magnitude varies by 4 magnitudes from the inner to the outer region, and
for most galaxies, this radial span corresponds to around 4 exponential scale
lengths. Since each scale length is one exponential factor and also about one
magnitude, the decline in luminosity is approximately proportional to the surface
density. This is the relationship for the Jeans mass given above. The variation
would be closer to 2 magnitudes if it were proportional to the square root of the
surface density in the case of noise limitations. Also in the noise interpretation,
the lack of bright cores in the outer regions would have to be explained for, e.g.
NGC 1566, where the upper envelope of the points has a trend with galactocentric
distance also.

The clumps in the main spiral arms of NGC 1566 get somewhat brighter with radius,
unlike the other distributions in Figure \ref{efremov_mag_radial}. For NGC 5194, the
arms have no obvious trend. The trend for NGC 1566 suggests that the disk thickness
in the arms increases with radius, possibly from a more constant velocity
dispersion. When the velocity dispersion is constant, both the thickness and the
Jeans mass vary as the inverse of the surface density \citep{inoue18}. This is
consistent with the trend we observe for the spiral arms in NGC 1566.

\subsection{A possible connection to the star formation rate}

\begin{figure} \epsscale{1.}
\includegraphics[width=6.in]{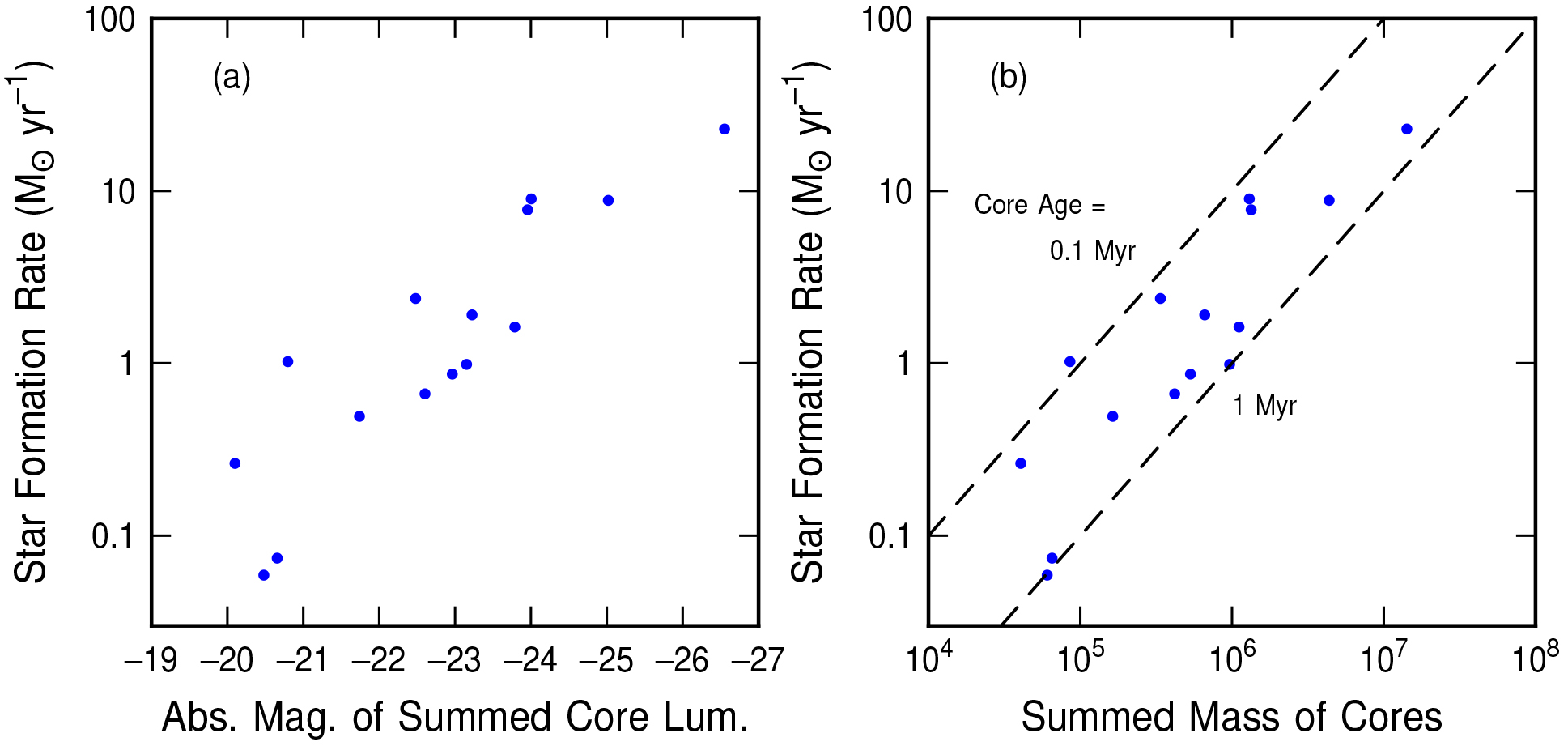}
\caption{Total galaxy star formation rates are plotted versus absolute magnitude
equivalents of the total $8\mu$m core luminosities (left) and the total core masses as determined
from their IRAC luminosities using a starburst spectral energy distribution and a population age
around 1 Myr (right).  }
\label{efremov_sfr}
\end{figure}

The total luminosity and stellar mass of the $8\mu$m clumps should be
proportional to the galaxy SFR if the clumps are the youngest stages of star
formation. Figure \ref{efremov_sfr} shows this is the case.  On the left, the SFRs
from Table 1 are plotted versus the summed $8\mu$m luminosities, converted to
absolute magnitudes (i.e., $-2.5\log_{10}$ of the sum over all cores of
$10^{-0.4[8.0]}$ for apparent magnitude $[8.0]$, converted to absolute magnitude by
the subtraction of $5\log_{10}(10^5D)$ for distance $D$ in Mpc).  Galaxies with
higher total SFRs have greater summed $8\mu$m luminosities of the cores. Note that
this is not the total $8\mu$m luminosity of the galaxy, but only the sum of the
relatively small core regions.

Figure \ref{efremov_sfr}(b) converts the IRAC luminosities into approximate stellar
masses using the procedure in Paper I. That is, the sum of the luminosities of the
four IRAC bands is multiplied by 10 to obtain an estimate of the total infrared
luminosity, based on integrals over SEDs for starbursts in \cite{xu01}. This total
clump IR luminosity is then converted to stellar mass of young stars using the
bolometric magnitude of a young stellar population in \cite{bruzual03}, which is
$-2.7944$ for solar metallicity at less than 1 Myr age. Using 4.74 mag as the
bolometric magnitude of the Sun, the bolometric luminosity per solar mass of young
stars is $3.01\times10^{35}\times10^{0.4*2.7944}$ at 1 Myr age. The inferred stellar
mass is the total IR luminosity of the clump divided by this quantity. The stellar
masses would be larger by a factor 5.1 for an age of 10 Myr.

Figure \ref{efremov_sfr}(b) shows, like Figure \ref{efremov_sfr}(a), the importance
of $8\mu$m clumps increasing with the SFR. With the clump mass, however, we can
divide the abscissa by the ordinate and get a time scale. This time would be the
average age of the clumps if all of the star formation in the galaxy came from them.
Considering the fainter clumps that are not included (cf. Fig. \ref{efremov_mags})
and the typical mass distribution for star forming regions, i.e., varying as the
inverse of mass for equal logarithmic intervals of mass, there could be an
approximately equal amount of clump mass in addition to what is plotted in Figure
\ref{efremov_sfr}(b), which would double the timescale. With this factor, the time
is between 0.2 Myr and 2 Myr, which is a reasonable age for a mostly-embedded young
stellar population. This result suggests that most of the star formation in these
galaxies goes through a phase when it is embedded in a dense $8\mu$ clump like what
we observe. This is an earlier phase than an HII region or FUV source, but the
H$\alpha$ or FUV luminosities can give about the same SFRs because they sample a
different time in the evolution of the clumps. The $24\mu$m luminosities presumably
sample both this embedded phase and a later, more exposed phase, but the $24\mu$m
images do not have the angular resolution to see the $3^{\prime\prime}$ diameter
cores that are visible at $8\mu$m.

\section{Conclusions}
\label{conc}

An unsharp mask technique has revealed several hundred small $8\mu$m cores in each
of 15 spiral galaxies observed with the Spitzer Space telescope. The cores have
colors suggestive of embedded star formation beneath $\sim15$ mag of visual
extinction combined with bright PAH emission. Nearly all spiral arms and spurs have
several IR cores, and more occur singly in smaller dust patches. Most cores do not
correspond to obvious optical features because of the extinction, and neither do
they stand out in the $8\mu$m or color Spitzer images because of confusion with
lower-intensity backgrounds.

The present study follows a more thorough view of similar cores in the galaxy M100,
published earlier \citep{elmegreen18}. The core colors there are about the same as
those here, as are the apparent magnitudes, which are set by selection effects. We
expect many more such cores will be found at higher resolution and greater
sensitivity. The usual luminosity function for young stellar clusters suggests that
all star-forming galaxies selected for angular size will have about the same number
and angular spacings for their $8\mu$m cores if observed to the same apparent
magnitude limit. That is the case here.

The morphological similarity between the observed galactic $8\mu$m cores with their
surrounding filaments and star-forming cores and filaments in the solar neighborhood
suggests a universal process of core formation based on gravitational collapse in
thin compressed regions.  For galaxies, the compression on the largest scale is from
stellar spiral wave activity. Because the interstellar medium on this large scale is
generally composed of smaller diffuse and self-gravitating clouds, the cores that
form would also involve some agglomeration of these smaller clouds, forced somewhat
by their mutual gravitational attraction. Other cores that are visible in $8\mu$m
rings in the outer parts of our galaxies also suggest gas compression and collapse
initiated by stellar feedback.

The positions of these cores inside spiral shock fronts suggests they are the first
step in the star formation process connected with spiral waves. When the young stars
emerge from their current high extinctions, they will presumably produce the usual
optical structures and age gradients that are seen in spiral arms.

The summed core luminosities and equivalent stellar masses correlate well with
the total galaxy star formation rates. The embedded stellar masses could account for
the star formation rate if the evolution phase of a typical core is between 0.2 Myr
and 2 Myr. This result suggests that most star formation goes through a dense
$8\mu$m core phase and that we are observing most of the young intermediate-to-high
mass star-formation sites in these galaxies.

Final note: This paper is dedicated to our friend and colleague for 27 years,
Dr. Yuri Nicolaevich Efremov (1937-2019), of the Sternberg Astronomical Observatory
in Moscow.

We are grateful to the referee for useful comments.

\newpage
\begin{deluxetable}{lccccccccc}
\tabletypesize{\scriptsize} \tablecolumns{10} \tablewidth{0pt}
\tablecaption{Galaxies} \tablehead{
\colhead{NGC}&
\colhead{Distance}&
\colhead{Distance}&
\colhead{No. of}&
\colhead{Hubble type}&
\colhead{Arm} &
\colhead{$R_{25}$} &
\colhead{$R_{25}$} &
\colhead{pc per}&
\colhead{SFR}\\
\colhead{}&
\colhead{(GSR), Mpc}&
\colhead{Modulus}&
\colhead{Cores}&
\colhead{}&
\colhead{Class}&
\colhead{arcmin}&
\colhead{kpc}&
\colhead{arcsec}&
\colhead{$M_\odot$ yr$^{-1}$}
}
\startdata
628	    &	11.13	&	30.23	&	681	&	SA(s)c I	    &	M	&	5.24	&	16.90	&	53.80 & 1.6	\\
1566	&	19.63	&	31.46	&	403	&	SAB(s)bc I    	&	G	&	4.16	&	23.67	&	94.88 & 8.8 	\\
2403	&	3.39	&	27.65	&	499	&	SAB(s)cd III	&	F	&	10.94	&	10.75	&	16.39 & 1.0 	\\
3184	&	8.8	    &	29.72	&	416	&	SAB(rs)cd II.2	&	M	&	3.71	&	9.46	&	42.53 & 0.074	\\
3351	&	9.98	&	30.00	&	324	&	SB(r)b II	    &	M	&	3.71	&	10.73	&	48.24 & 0.66	\\
3938	&	12.52	&	30.49	&	353	&	SA(s)c I     	&	M	&	2.69	&	9.75	&	60.51 & 0.87	\\
4254	&	34.73	&	32.70	&	469	&	SA(s)c I.3	    &	M	&	2.69	&	27.04	&	167.86 & 22.8	\\
4579	&	21.62	&	31.67	&	179	&	SAB(rs)b II    	&	M	&	2.94	&	18.46	&	104.50 & 1.9\\
4625	&	9.88	&	29.97	&	69	&	SAB(rs)m pec V	&	F	&	1.09	&	3.13	&	47.75 & 0.059	\\
4725	&	17.8	&	31.25	&	272	&	SAB(r)ab pec II	&	M	&	5.36	&	27.66	&	86.03 & 0.98	\\
4736	&	5.31	&	28.63	&	169	&	(R)SA(r)ab I-II	&	F	&	5.61	&	8.64	&	25.67 & 0.49	\\
5194	&	8.12	&	29.55	&	670	&	SA(s)bc pec I-II&	G	&	5.61	&	13.21	&	39.25 & 7.8	\\
5236	&	5.63	&	28.75	&	857	&	SAB(s)c II	    &	M	&	6.44	&	10.52	&	27.21 & 9.0	\\
6946	&	3.93	&	27.97	&	725	&	SAB(rs)cd II	&	M	&	5.74	&	6.54	&	19.00 & 2.4	\\
7793	&	3.38	&	27.65	&	229	&	SA(s)d IV	    &	F	&	4.67	&	4.57    &	16.34  & 0.26
\enddata
\tablenotetext{1}{Star formation rates are from H$\alpha$ and $24\mu$m in
\cite{kennicutt11} in all cases but NGC 1566, NGC 2403, NGC 5194, and NGC 5236,
where they are from FUV and total IR in \cite{thilker07}.}
\label{galaxylist}
\end{deluxetable}

\newpage
\begin{deluxetable}{llccccccc}
\tabletypesize{\scriptsize} \tablecolumns{9} \tablewidth{0pt} \tablecaption{$8\mu$m
Cores in NGC 628} \tablehead{ \colhead{RA}& \colhead{DEC}& \colhead{[3.6]}&
\colhead{[4.5]}& \colhead{[5.8]} & \colhead{[8.0]} & \colhead{[3.6]-[4.5]} &
\colhead{[4.5]-[5.8]} &
\colhead{[5.8]-[8.0]} \\
} \startdata
 1:36:59.96& 15:46:14.3&$  19.82\pm   2.24$&$  19.11\pm   1.97$&$  16.58\pm   0.78$&$  15.03\pm   0.53$&   0.71&   2.52&   1.55\\
 1:36:59.76& 15:46:17.0&$  18.76\pm   1.34$&$  18.69\pm   1.62$&$  20.46\pm   6.34$&$  15.39\pm   0.65$&   0.07&  -1.77&   5.07\\
 1:36:59.29& 15:46:17.3&$  17.25\pm   0.67$&$  16.96\pm   0.73$&$  14.84\pm   0.34$&$  13.20\pm   0.22$&   0.29&   2.13&   1.64\\
 1:36:59.25& 15:46:23.7&$  17.80\pm   0.87$&$  17.51\pm   0.94$&$  15.96\pm   0.58$&$  14.36\pm   0.38$&   0.30&   1.54&   1.60\\
 1:36:58.96& 15:46:57.8&$  18.81\pm   1.37$&$  19.38\pm   2.22$&$  17.30\pm   1.09$&$  15.26\pm   0.56$&  -0.57&   2.08&   2.04\\
...
\enddata
\label{listofclumps628}
\end{deluxetable}
\newpage

\begin{deluxetable}{llccccccc}
\tabletypesize{\scriptsize} \tablecolumns{9} \tablewidth{0pt} \tablecaption{$8\mu$m
Cores in NGC 1566} \tablehead{ \colhead{RA}& \colhead{DEC}& \colhead{[3.6]}&
\colhead{[4.5]}& \colhead{[5.8]} & \colhead{[8.0]} & \colhead{[3.6]-[4.5]} &
\colhead{[4.5]-[5.8]} &
\colhead{[5.8]-[8.0]} \\
} \startdata
 4:20:28.68&-54:56: 5.9&$  17.62\pm   0.79$&$  17.05\pm   0.76$&$  17.43\pm   1.13$&$  15.22\pm   0.55$&   0.57&  -0.38&   2.21\\
 4:20:26.54&-54:54:47.6&$  19.06\pm   1.54$&$  19.13\pm   1.98$&$  17.14\pm   0.99$&$  15.40\pm   0.60$&  -0.07&   1.99&   1.75\\
 4:20:21.27&-54:57:19.2&$  15.82\pm   0.35$&$  15.72\pm   0.41$&$  16.16\pm   0.63$&$  15.32\pm   0.58$&   0.10&  -0.44&   0.84\\
 4:20:20.05&-54:54: 2.7&$  15.98\pm   0.37$&$  15.73\pm   0.41$&$  16.32\pm   0.67$&$  14.70\pm   0.43$&   0.24&  -0.58&   1.62\\
 4:20:18.43&-54:53:43.3&$  19.36\pm   1.77$&$  19.05\pm   1.90$&$  17.65\pm   1.25$&$  15.40\pm   0.60$&   0.31&   1.40&   2.25\\
...
\enddata
\label{listofclumps1566}
\end{deluxetable}
\newpage

\begin{deluxetable}{llccccccc}
\tabletypesize{\scriptsize} \tablecolumns{9} \tablewidth{0pt} \tablecaption{$8\mu$m
Cores in NGC 2403} \tablehead{ \colhead{RA}& \colhead{DEC}& \colhead{[3.6]}&
\colhead{[4.5]}& \colhead{[5.8]} & \colhead{[8.0]} & \colhead{[3.6]-[4.5]} &
\colhead{[4.5]-[5.8]} &
\colhead{[5.8]-[8.0]} \\
} \startdata
  7:37:51.91& 65:34:17.8&$  17.72\pm   0.83$&$  17.08\pm   0.77$&$  15.80\pm   0.53$&$  14.23\pm   0.35$&   0.64&   1.28&   1.57\\
 7:37:49.88& 65:31:38.9&$  19.17\pm   1.63$&$  18.99\pm   1.86$&$  17.05\pm   0.95$&$  15.51\pm   0.63$&   0.18&   1.94&   1.54\\
 7:37:49.39& 65:31:29.1&$  16.79\pm   0.54$&$  16.35\pm   0.55$&$  16.48\pm   0.73$&$  15.09\pm   0.52$&   0.44&  -0.13&   1.39\\
 7:37:46.66& 65:32:52.4&$  20.05\pm   2.45$&$  19.35\pm   2.19$&$  17.57\pm   1.21$&$  15.77\pm   0.72$&   0.70&   1.78&   1.80\\
 7:37:44.25& 65:32:57.0&$  20.16\pm   2.61$&$  17.88\pm   1.11$&$  16.91\pm   0.88$&$  15.76\pm   0.71$&   2.28&   0.97&   1.14\\
...
\enddata
\label{listofclumps2403}
\end{deluxetable}
\newpage
\begin{deluxetable}{llccccccc}
\tabletypesize{\scriptsize} \tablecolumns{9} \tablewidth{0pt} \tablecaption{$8\mu$m
Cores in NGC 3184} \tablehead{ \colhead{RA}& \colhead{DEC}& \colhead{[3.6]}&
\colhead{[4.5]}& \colhead{[5.8]} & \colhead{[8.0]} & \colhead{[3.6]-[4.5]} &
\colhead{[4.5]-[5.8]} &
\colhead{[5.8]-[8.0]} \\
} \startdata
10:18:32.52& 41:26:28.1&$  19.08\pm   1.55$&$  18.77\pm   1.68$&$  16.66\pm   0.80$&$  15.14\pm   0.53$&   0.30&   2.12&   1.52\\
10:18:31.52& 41:26:38.2&$  18.14\pm   1.01$&$  17.38\pm   0.88$&$  15.87\pm   0.55$&$  14.67\pm   0.44$&   0.76&   1.50&   1.21\\
10:18:31.39& 41:26:42.7&$  18.14\pm   1.01$&$  17.93\pm   1.14$&$  16.10\pm   0.62$&$  14.34\pm   0.38$&   0.21&   1.82&   1.76\\
10:18:31.22& 41:26:52.8&$  18.59\pm   1.24$&$  17.99\pm   1.17$&$  17.17\pm   1.03$&$  15.90\pm   0.76$&   0.60&   0.82&   1.27\\
10:18:31.06& 41:26:42.7&$  17.31\pm   0.69$&$  17.09\pm   0.78$&$  15.93\pm   0.60$&$  14.43\pm   0.41$&   0.22&   1.15&   1.50\\
...
\enddata
\label{listofclumps3184}
\end{deluxetable}
\newpage
\begin{deluxetable}{llccccccc}
\tabletypesize{\scriptsize} \tablecolumns{9} \tablewidth{0pt} \tablecaption{$8\mu$m
Cores in NGC 3351} \tablehead{ \colhead{RA}& \colhead{DEC}& \colhead{[3.6]}&
\colhead{[4.5]}& \colhead{[5.8]} & \colhead{[8.0]} & \colhead{[3.6]-[4.5]} &
\colhead{[4.5]-[5.8]} &
\colhead{[5.8]-[8.0]} \\
} \startdata
10:44: 7.01& 11:41:16.2&$  19.65\pm   2.02$&$  19.44\pm   2.29$&$  17.32\pm   1.09$&$  15.25\pm   0.56$&   0.21&   2.13&   2.07\\
10:44: 6.13& 11:43:51.6&$  19.67\pm   2.05$&$  19.32\pm   2.15$&$  16.70\pm   0.81$&$  15.35\pm   0.59$&   0.36&   2.62&   1.35\\
10:44: 6.00& 11:43:14.1&$  20.04\pm   2.42$&$  20.86\pm   4.44$&$  17.94\pm   1.46$&$  15.49\pm   0.65$&  -0.82&   2.92&   2.45\\
10:44: 5.86& 11:41:16.6&$  19.88\pm   2.27$&$  19.90\pm   2.85$&$  18.07\pm   1.53$&$  15.90\pm   0.76$&  -0.02&   1.83&   2.17\\
10:44: 5.85& 11:41:50.1&$  21.04\pm   3.94$&$  19.29\pm   2.13$&$  17.12\pm   1.02$&$  15.42\pm   0.62$&   1.75&   2.17&   1.70\\
...
\enddata
\label{listofclumps3351}
\end{deluxetable}
\newpage

\begin{deluxetable}{llccccccc}
\tabletypesize{\scriptsize} \tablecolumns{9} \tablewidth{0pt} \tablecaption{$8\mu$m
Cores in NGC 3938} \tablehead{ \colhead{RA}& \colhead{DEC}& \colhead{[3.6]}&
\colhead{[4.5]}& \colhead{[5.8]} & \colhead{[8.0]} & \colhead{[3.6]-[4.5]} &
\colhead{[4.5]-[5.8]} &
\colhead{[5.8]-[8.0]} \\
} \startdata
11:53: 2.06& 44: 7:20.9&$  19.56\pm   1.95$&$  18.87\pm   1.76$&$  17.24\pm   1.06$&$  15.44\pm   0.61$&   0.68&   1.63&   1.81\\
11:53: 0.26& 44: 7:28.3&$  19.00\pm   1.50$&$  18.80\pm   1.70$&$  17.32\pm   1.09$&$  15.48\pm   0.63$&   0.20&   1.48&   1.84\\
11:53: 0.26& 44: 7:48.4&$  15.72\pm   0.33$&$  14.94\pm   0.29$&$  13.59\pm   0.19$&$  11.96\pm   0.13$&   0.78&   1.35&   1.63\\
11:53: 0.10& 44: 7: 0.2&$  18.07\pm   0.98$&$  17.59\pm   0.98$&$  16.28\pm   0.66$&$  14.64\pm   0.43$&   0.48&   1.31&   1.64\\
11:53: 0.02& 44: 8: 0.0&$  16.60\pm   0.50$&$  16.14\pm   0.50$&$  14.67\pm   0.32$&$  12.84\pm   0.19$&   0.46&   1.47&   1.82\\
...
\enddata
\label{listofclumps3938}
\end{deluxetable}
\newpage

\begin{deluxetable}{llccccccc}
\tabletypesize{\scriptsize} \tablecolumns{9} \tablewidth{0pt} \tablecaption{$8\mu$m
Cores in NGC 4254} \tablehead{ \colhead{RA}& \colhead{DEC}& \colhead{[3.6]}&
\colhead{[4.5]}& \colhead{[5.8]} & \colhead{[8.0]} & \colhead{[3.6]-[4.5]} &
\colhead{[4.5]-[5.8]} &
\colhead{[5.8]-[8.0]} \\
} \startdata
12:18:58.38& 14:26:19.3&$  19.81\pm   2.18$&$  19.89\pm   2.81$&$  18.48\pm   1.91$&$  15.97\pm   0.79$&  -0.08&   1.41&   2.51\\
12:18:57.71& 14:26:48.9&$  18.52\pm   1.21$&$  17.98\pm   1.17$&$  16.07\pm   0.62$&$  14.23\pm   0.37$&   0.54&   1.91&   1.84\\
12:18:57.66& 14:25: 5.0&$  19.33\pm   1.75$&$  19.23\pm   2.07$&$  17.45\pm   1.20$&$  15.24\pm   0.57$&   0.10&   1.78&   2.21\\
12:18:57.52& 14:25:18.3&$  14.23\pm   0.17$&$  14.32\pm   0.22$&$  14.10\pm   0.26$&$  12.95\pm   0.28$&  -0.09&   0.22&   1.15\\
12:18:57.45& 14:25:31.9&$  18.65\pm   1.28$&$  18.44\pm   1.44$&$  16.17\pm   0.65$&$  14.33\pm   0.39$&   0.21&   2.27&   1.84\\
...
\enddata
\label{listofclumps4254}
\end{deluxetable}
\newpage

\begin{deluxetable}{llccccccc}
\tabletypesize{\scriptsize} \tablecolumns{9} \tablewidth{0pt} \tablecaption{$8\mu$m
Cores in NGC 4579} \tablehead{ \colhead{RA}& \colhead{DEC}& \colhead{[3.6]}&
\colhead{[4.5]}& \colhead{[5.8]} & \colhead{[8.0]} & \colhead{[3.6]-[4.5]} &
\colhead{[4.5]-[5.8]} &
\colhead{[5.8]-[8.0]} \\
} \startdata
12:37:50.13& 11:48:38.5&$  19.69\pm   2.11$&$  19.18\pm   2.02$&$  19.81\pm   5.10$&$  15.32\pm   0.60$&   0.51&  -0.63&   4.49\\
12:37:49.81& 11:48:37.4&$  18.60\pm   1.27$&$  17.83\pm   1.09$&$  17.06\pm   0.96$&$  15.09\pm   0.53$&   0.78&   0.77&   1.97\\
12:37:49.25& 11:48:46.0&$  19.75\pm   2.13$&$  20.10\pm   3.15$&$  18.73\pm   2.19$&$  15.65\pm   0.73$&  -0.36&   1.37&   3.08\\
12:37:49.25& 11:49:26.5&$  19.26\pm   1.77$&$  18.81\pm   1.73$&$  16.80\pm   0.86$&$  14.81\pm   0.50$&   0.45&   2.01&   1.99\\
12:37:49.17& 11:48:20.5&$  18.92\pm   1.47$&$  18.58\pm   1.55$&$  16.60\pm   0.78$&$  15.04\pm   0.52$&   0.34&   1.99&   1.55\\
...
\enddata
\label{listofclumps4579}
\end{deluxetable}
\newpage

\begin{deluxetable}{llccccccc}
\tabletypesize{\scriptsize} \tablecolumns{9} \tablewidth{0pt} \tablecaption{$8\mu$m
Cores in NGC 4625} \tablehead{ \colhead{RA}& \colhead{DEC}& \colhead{[3.6]}&
\colhead{[4.5]}& \colhead{[5.8]} & \colhead{[8.0]} & \colhead{[3.6]-[4.5]} &
\colhead{[4.5]-[5.8]} &
\colhead{[5.8]-[8.0]} \\
} \startdata
12:41:54.82& 41:16:18.5&$  18.29\pm   1.11$&$  17.89\pm   1.13$&$  15.93\pm   0.58$&$  14.21\pm   0.37$&   0.40&   1.96&   1.72\\
12:41:54.82& 41:16: 0.6&$  17.92\pm   0.91$&$  17.63\pm   1.00$&$  15.67\pm   0.50$&$  13.93\pm   0.31$&   0.28&   1.97&   1.74\\
12:41:54.56& 41:16:19.0&$  18.84\pm   1.40$&$  18.23\pm   1.30$&$  16.25\pm   0.67$&$  14.54\pm   0.42$&   0.61&   1.97&   1.71\\
12:41:54.54& 41:15:58.9&$  16.93\pm   0.58$&$  16.65\pm   0.63$&$  14.59\pm   0.31$&$  12.86\pm   0.19$&   0.28&   2.07&   1.72\\
12:41:54.50& 41:16:10.3&$  17.24\pm   0.67$&$  17.06\pm   0.76$&$  15.19\pm   0.40$&$  13.51\pm   0.25$&   0.18&   1.87&   1.68\\
...
\enddata
\label{listofclumps4625}
\end{deluxetable}
\newpage

\begin{deluxetable}{llccccccc}
\tabletypesize{\scriptsize} \tablecolumns{9} \tablewidth{0pt} \tablecaption{$8\mu$m
Cores in NGC 4725} \tablehead{ \colhead{RA}& \colhead{DEC}& \colhead{[3.6]}&
\colhead{[4.5]}& \colhead{[5.8]} & \colhead{[8.0]} & \colhead{[3.6]-[4.5]} &
\colhead{[4.5]-[5.8]} &
\colhead{[5.8]-[8.0]} \\
} \startdata
12:50:38.39& 25:32:19.5&$  19.23\pm   1.66$&$  20.03\pm   3.02$&$  17.59\pm   1.25$&$  15.46\pm   0.62$&  -0.81&   2.44&   2.13\\
12:50:36.65& 25:32:32.1&$  19.47\pm   1.86$&$  19.55\pm   2.41$&$  18.03\pm   1.49$&$  15.78\pm   0.73$&  -0.09&   1.52&   2.25\\
12:50:36.19& 25:33:14.0&$  19.01\pm   1.51$&$  18.55\pm   1.51$&$  16.42\pm   0.72$&$  14.68\pm   0.46$&   0.46&   2.13&   1.74\\
12:50:36.12& 25:33:33.7&$  19.08\pm   1.55$&$  18.89\pm   1.77$&$  17.60\pm   1.24$&$  15.30\pm   0.57$&   0.19&   1.29&   2.30\\
12:50:35.97& 25:33:14.0&$  18.87\pm   1.41$&$  18.98\pm   1.85$&$  16.43\pm   0.73$&$  14.56\pm   0.41$&  -0.11&   2.55&   1.87\\
...
\enddata
\label{listofclumps4725}
\end{deluxetable}
\newpage

\begin{deluxetable}{llccccccc}
\tabletypesize{\scriptsize} \tablecolumns{9} \tablewidth{0pt} \tablecaption{$8\mu$m
Cores in NGC 4736} \tablehead{ \colhead{RA}& \colhead{DEC}& \colhead{[3.6]}&
\colhead{[4.5]}& \colhead{[5.8]} & \colhead{[8.0]} & \colhead{[3.6]-[4.5]} &
\colhead{[4.5]-[5.8]} &
\colhead{[5.8]-[8.0]} \\
} \startdata
12:51: 9.80& 41: 6:10.2&$  19.67\pm   2.05$&$  19.34\pm   2.18$&$  17.55\pm   1.23$&$  15.51\pm   0.66$&   0.33&   1.78&   2.04\\
12:51: 9.73& 41: 6:15.1&$  18.92\pm   1.45$&$  18.56\pm   1.52$&$  16.77\pm   0.85$&$  15.06\pm   0.52$&   0.37&   1.78&   1.72\\
12:51: 6.27& 41: 7:45.9&$  17.90\pm   0.91$&$  17.73\pm   1.04$&$  16.08\pm   0.61$&$  14.18\pm   0.34$&   0.17&   1.64&   1.90\\
12:51: 6.12& 41: 7: 6.7&$  18.40\pm   1.13$&$  18.31\pm   1.36$&$  18.20\pm   1.70$&$  15.69\pm   0.71$&   0.09&   0.11&   2.51\\
12:51: 6.12& 41: 6:50.9&$  18.83\pm   1.39$&$  18.65\pm   1.58$&$  17.70\pm   1.29$&$  15.57\pm   0.66$&   0.18&   0.95&   2.13\\
...
\enddata
\label{listofclumps4736}
\end{deluxetable}
\newpage

\begin{deluxetable}{llccccccc}
\tabletypesize{\scriptsize} \tablecolumns{9} \tablewidth{0pt} \tablecaption{$8\mu$m
Cores in NGC 5194} \tablehead{ \colhead{RA}& \colhead{DEC}& \colhead{[3.6]}&
\colhead{[4.5]}& \colhead{[5.8]} & \colhead{[8.0]} & \colhead{[3.6]-[4.5]} &
\colhead{[4.5]-[5.8]} &
\colhead{[5.8]-[8.0]} \\
} \startdata
13:30: 8.58& 47:13:36.6&$  17.70\pm   0.83$&$  17.64\pm   1.00$&$  16.00\pm   0.60$&$  14.46\pm   0.44$&   0.06&   1.63&   1.54\\
13:30: 8.42& 47:12:26.8&$  18.78\pm   1.40$&$  18.50\pm   1.51$&$  16.09\pm   0.66$&$  14.05\pm   0.35$&   0.28&   2.42&   2.04\\
13:30: 8.42& 47:12:14.8&$  18.09\pm   1.00$&$  17.61\pm   0.98$&$  15.93\pm   0.58$&$  14.24\pm   0.42$&   0.48&   1.67&   1.70\\
13:30: 8.36& 47:13:41.1&$  18.25\pm   1.10$&$  17.88\pm   1.12$&$  15.38\pm   0.47$&$  13.58\pm   0.32$&   0.37&   2.50&   1.81\\
13:30: 8.28& 47:12:40.3&$  18.25\pm   1.08$&$  18.05\pm   1.22$&$  15.33\pm   0.45$&$  13.88\pm   0.39$&   0.20&   2.72&   1.45\\
...
\enddata
\label{listofclumps5194}
\end{deluxetable}
\newpage

\begin{deluxetable}{llccccccc}
\tabletypesize{\scriptsize} \tablecolumns{9} \tablewidth{0pt} \tablecaption{$8\mu$m
Cores in NGC 5236} \tablehead{ \colhead{RA}& \colhead{DEC}& \colhead{[3.6]}&
\colhead{[4.5]}& \colhead{[5.8]} & \colhead{[8.0]} & \colhead{[3.6]-[4.5]} &
\colhead{[4.5]-[5.8]} &
\colhead{[5.8]-[8.0]} \\
} \startdata
13:37:19.27&-29:52:21.8&$  19.56\pm   1.96$&$  19.36\pm   2.21$&$  18.18\pm   1.71$&$  15.46\pm   0.65$&   0.21&   1.18&   2.72\\
13:37:19.15&-29:52:14.0&$  17.42\pm   0.73$&$  17.30\pm   0.85$&$  16.82\pm   0.86$&$  15.01\pm   0.51$&   0.12&   0.48&   1.81\\
13:37:18.89&-29:52:50.4&$  18.95\pm   1.48$&$  19.27\pm   2.12$&$  17.03\pm   0.94$&$  15.33\pm   0.60$&  -0.32&   2.24&   1.70\\
13:37:18.46&-29:52:42.5&$  25.60\pm  99.00$&$  19.79\pm   2.70$&$  16.77\pm   0.83$&$  14.80\pm   0.46$&   5.80&   3.03&   1.97\\
13:37:18.44&-29:52: 0.1&$  18.72\pm   1.35$&$  19.02\pm   1.89$&$  15.72\pm   0.52$&$  13.92\pm   0.32$&  -0.30&   3.30&   1.80\\
...
\enddata
\label{listofclumps5236}
\end{deluxetable}
\newpage

\begin{deluxetable}{llccccccc}
\tabletypesize{\scriptsize} \tablecolumns{9} \tablewidth{0pt} \tablecaption{$8\mu$m
Cores in NGC 6946} \tablehead{ \colhead{RA}& \colhead{DEC}& \colhead{[3.6]}&
\colhead{[4.5]}& \colhead{[5.8]} & \colhead{[8.0]} & \colhead{[3.6]-[4.5]} &
\colhead{[4.5]-[5.8]} &
\colhead{[5.8]-[8.0]} \\
} \startdata
20:35:26.74& 60: 8:50.3&$  19.30\pm   1.73$&$  18.77\pm   1.68$&$  17.55\pm   1.20$&$  15.77\pm   0.72$&   0.53&   1.22&   1.77\\
20:35:26.65& 60: 9: 9.8&$  20.70\pm   3.34$&$  18.79\pm   1.71$&$  16.85\pm   0.86$&$  15.33\pm   0.60$&   1.90&   1.94&   1.52\\
20:35:26.65& 60: 9:13.6&$  19.47\pm   1.87$&$  19.42\pm   2.29$&$  16.87\pm   0.90$&$  14.81\pm   0.47$&   0.05&   2.56&   2.05\\
20:35:26.34& 60: 8:42.8&$  19.40\pm   1.84$&$  18.63\pm   1.58$&$  16.24\pm   0.66$&$  14.90\pm   0.50$&   0.77&   2.39&   1.34\\
20:35:26.36& 60:10:10.6&$  16.23\pm   0.42$&$  15.85\pm   0.44$&$  14.65\pm   0.33$&$  13.15\pm   0.24$&   0.38&   1.20&   1.50\\
...
\enddata
\label{listofclumps6946}
\end{deluxetable}
\newpage

\begin{deluxetable}{llccccccc}
\tabletypesize{\scriptsize} \tablecolumns{9} \tablewidth{0pt} \tablecaption{$8\mu$m
Cores in NGC 7793} \tablehead{ \colhead{RA}& \colhead{DEC}& \colhead{[3.6]}&
\colhead{[4.5]}& \colhead{[5.8]} & \colhead{[8.0]} & \colhead{[3.6]-[4.5]} &
\colhead{[4.5]-[5.8]} &
\colhead{[5.8]-[8.0]} \\
} \startdata
23:58:14.20&-32:36:35.4&$  19.22\pm   1.66$&$  19.19\pm   2.04$&$  17.57\pm   1.23$&$  15.68\pm   0.69$&   0.03&   1.62&   1.89\\
23:58:12.26&-32:36:38.7&$  18.07\pm   0.98$&$  17.92\pm   1.13$&$  16.62\pm   0.77$&$  15.67\pm   0.69$&   0.15&   1.31&   0.95\\
23:58:11.95&-32:36:40.7&$  19.52\pm   1.90$&$  19.46\pm   2.30$&$  17.24\pm   1.05$&$  15.65\pm   0.68$&   0.06&   2.22&   1.59\\
23:58: 8.82&-32:36:49.1&$  17.88\pm   0.90$&$  17.17\pm   0.80$&$  16.36\pm   0.70$&$  15.17\pm   0.55$&   0.70&   0.82&   1.18\\
23:58: 1.27&-32:36:45.5&$  18.52\pm   1.20$&$  18.81\pm   1.71$&$  16.25\pm   0.67$&$  14.53\pm   0.42$&  -0.28&   2.55&   1.72\\
...
\enddata
\label{listofclumps7793}
\end{deluxetable}
\newpage


\begin{thebibliography}

\bibitem[Allen(2004)]{allen04} Allen, L., Calvet, N., D'Alessio, P., Merin, B.,
    Hartmann, L., et al. 2004, ApJS, 154, 363

\bibitem[Andr\'e(2017)]{andre17} Andr\'e, P. 2017, CRGeo, 349, 187

\bibitem[Baba et al.(2013)]{baba13} Baba, J., Saitoh, T.R., \& Wada,
    K. 2013, ApJ, 763, 46

\bibitem[Balbus(1988)]{balbus88} Balbus, S.A. 1988, ApJ, 324, 60

\bibitem[Balbus \& Cowie(1985)]{balbus85} Balbus, S. A., \& Cowie,
    L. L. 1985, ApJ, 297, 61

\bibitem[Block(1984)]{block84} Block, D. L. 1984, in A photographic
    atlas of primarily late type spirals printed as if each galaxy were at the same
    distance, Fort Hare University Press, South Africa. 36 pp.

\bibitem[Bohlin et al.(1978)]{bohlin78} Bohlin, R. C., Savage, B. D., \& Drake, J.
    F. 1978, ApJ, 224, 132

\bibitem[Bruzual \& Charlot(2003)]{bruzual03} Bruzual, G.,  \& Charlot, S. 2003,
    MNRAS, 344, 1000

\bibitem[Chakrabarti(2003)]{chakrabarti03} Chakrabarti S., Laughlin G., \& Shu F.
    H., 2003, ApJ, 596, 220

\bibitem[Chandrasekhar \& Fermi(1953)]{chandra53} Chandrasekhar, S. \& Fermi, E.
    1953, ApJ, 118, 116

\bibitem[Chira et al.(2018)]{chira18} Chira, R. -A., Kainulainen, J.,
    Ib\'a\~nez-Mej\'ia, J. C., Henning, Th., Mac Low, M. -M. 2018, A\&A, 610, A62

\bibitem[Chou et al.(2000)]{chou00} Chou, W., Matsumoto, R., Tajima, T., Umekawa,
    M., \& Shibata, K. 2000, ApJ, 538, 710

\bibitem[Clarke et al.(2017)]{clarke17} Clarke, S. D., Whitworth, A. P.,
    Duarte-Cabral, A., \& Hubber, D. A. 2017, MNRAS, 468, 2489

\bibitem[Cowie(1981)]{cowie81} Cowie, L. L. 1981, ApJ, 245, 66

\bibitem[Dobbs \& Bonnell(2008)]{dobbs08a} Dobbs, C.L. \& Bonnell, I.A. 2008a,
    MNRAS, 385, 1893

\bibitem[Dobbs(2008)]{dobbs08} Dobbs, C.L. 2008, MNRAS, 391, 844

\bibitem[Dobbs \& Pringle(2010)]{dobbs10} Dobbs C. L. \& Pringle J. E., 2010, MNRAS,
    409, 396

\bibitem[Dobbs \& Baba(2014)]{dobbs14} Dobbs C. \& Baba J. 2014, PASA, 31, 35

\bibitem[Efremov(2010)]{efremov10} Efremov, Yu. N. 2010, MNRAS, 405, 1531

\bibitem[Elmegreen(1979)]{elmegreen79} Elmegreen, B.G. 1979, ApJ, 231, 372

\bibitem[Elmegreen(1982)]{elmegreen82} Elmegreen, B.G. 1982. ApJ, 253, 634

\bibitem[Elmegreen \& Elmegreen(1983)]{elmegreen83} Elmegreen, B.G., \& Elmegreen,
    D.M. 1983, MNRAS, 203, 31

\bibitem[Elmegreen \& Elmegreen(1986)]{elmegreen86} Elmegreen, B.G., \& Elmegreen,
    D.M. 1986, ApJ, 311, 554

\bibitem[Elmegreen \& Elmegreen(1987)]{elmegreen87} Elmegreen, B.G. \&  Elmegreen,
    D.M. 1987, ApJ, 320, 182

\bibitem[Elmegreen \& Elmegreen(1987b)]{elmegreen87b} Elmegreen, D.M. \&  Elmegreen,
    B.G. 1987, ApJ, 314, 3

\bibitem[Elmegreen(1994)]{elmegreen94} Elmegreen, B.G. 1994, ApJ, 433, 39

\bibitem[Elmegreen et al.(2018)]{elmegreen18} Elmegreen, B.G., Elmegreen, D.M., \&
    Efremov, Y.N 2018, ApJ, 863, 59

\bibitem[Elmegreen et al.(2006)]{elmegreen06} Elmegreen, D.M., Elmegreen, B.G.,
    Kaufman, M., et al. 2006, ApJ, 642, 158

\bibitem[Fiege \& Pudritz(2000)]{fiege00} Fiege, J.D. \& Pudritz, R.E. 2000, MNRAS,
    311, 105

\bibitem[Goldreich \& Lynden Bell(1965)]{goldreich65} Goldreich, P., \& Lynden-Bell,
    D. 1965, MNRAS, 130, 97

\bibitem[Goodman et al.(2014)]{goodman14} Goodman, A.A., Alves, J., Beaumont, C.N.,
    et al. 2014, ApJ, 797, 53

\bibitem[Grabelsky et al.(1987)]{grabelsky87} Grabelsky, D. A., Cohen, R. S.,
    Bronfman, L., Thaddeus, P., \& May, J. 1987, ApJ, 315, 122

\bibitem[Gusev \& Efremov(2013)]{gusev13} Gusev, A. S., \& Efremov, Yu. N. 2013,
    MNRAS, 434, 313

\bibitem[Gutermuth et al.(2009)]{gutermuth09} Gutermuth, R. A., Megeath, S. T.,
    Myers, P. C., et al. 2009, ApJS, 184, 18

\bibitem[Hosseinirad et al.(2018)]{hoss18} Hosseinirad,M., Abbassi, S., Roshan, M.,
    Naficy, K. 2018, MNRAS, 475, 2632

\bibitem[Inoue \& Yoshida(2018)]{inoue18} Inoue, S. \& Yoshida, N. 2018, MNRAS, 474,
    3466

\bibitem[Inoue \& Yoshida(2019)]{inoue19} Inoue, S. \& Yoshida, N. 2019, MNRAS, 485,
    3024

\bibitem[Inutsuka \& Miyama(1992)]{inutsuka92} Inutsuka, S. \& Miyama, S.M. 1992,
    ApJ, 388, 392

\bibitem[Kainulainen et al.(2016)]{kainulainen16} Kainulainen, J., Hacar, A., Alves,
    J., Beuther, H., Bouy, H., \& Tafalla, M. 2016, A\&A, 586, A27

\bibitem[Kennicutt et al.(2003)]{kennicutt03} Kennicutt, R.C., Jr., Armus, L.,
    Bendo, G. et al. 2003, PASP, 115, 928

\bibitem[Kennicutt et al.(2011)]{kennicutt11} Kennicutt, R.C., Calzetti, D., Aniano
    G., et al. 2011, PASP, 123, 1347

\bibitem[Khoperskov et al.(2013)]{khoperskov13} Khoperskov S. A., Vasiliev E. O.,
    Sobolev A. M., \& Khoperskov A. V., 2013, MNRAS, 428, 2311

\bibitem[Kim \& Ostriker(2001)]{kim01} Kim, W.-T., \& Ostriker, E.C. 2001, ApJ, 559,
    70

\bibitem[Kim et al.(2002a)]{kim02a} Kim, W.-T., Ostriker, E.C., \& Stone, J.M. 2002,
    ApJ, 581, 1080

\bibitem[Kim \& Ostriker(2002b)]{kim02b} Kim, W.-T., \& Ostriker, E.C. 2002b, ApJ,
    570, 132

\bibitem[Kim \& Ostriker(2006)]{kim06} Kim, T.-T., \& Ostriker, E.C. 2006, ApJ, 646,
    213

\bibitem[Kim \& Ostriker(2007)]{kim07} Kim, W.-T., \& Ostriker, E.C. 2007, ApJ, 660,
    1232

\bibitem[Koo et al.(2018)]{koo18} Koo, B.-C., Park, G., Kim, W.-T., Lee, M.G.,
    Balser, D., Wenger, T. 2018, AAS, 23123704

\bibitem[Kratter \& Lodato(2016)]{kratter16} Kratter, K., \& Lodato, G. 2016,
    ARA\&A, 54, 271

\bibitem[Kwan \& Valdes(1983)]{kwan83} Kwan, J. \& Valdes, F. 1983, ApJ, 271, 604

\bibitem[Lee \& Shu(2012)]{lee12} Lee, W.-K., \& Shu, F.H. 2012, ApJ, 756, 45

\bibitem[Li \& Draine(2001)]{li01} Li, A., \& Draine, B.T. 2001, ApJ, 554, 778

\bibitem[Lou \& Hu(2017)]{lou17} Lou, Y.-Q. \& Hu, X.Y. 2017, MNRAS, 468, 2771

\bibitem[Mart\'inez-Garc\'ia et al.(2009)]{martinez09} Mart\'inez-Garc\'ia, E. E.,
    Gonz\'alez-L\'opezlira, R.A., Bruzual, A.G., 2009, ApJ, 694, 512

\bibitem[Mattern et al.(2018)]{mattern18} Mattern, M., Kainulainen, J., Zhang, M.,
    \& Beuther, H. 2018, A\&A, 616, A78

\bibitem[Megeath et al.(2004)]{megeath04} Megeath, S.T., Allen, L.E., Gutermuth,
    R.A., Pipher, J. L., Myers, P.C., Calvet, N., Hartmann, L., Muzerolle, J., \&
    Fazio, G. G.  2004, ApJS, 154, 367

\bibitem[McGee \& Milton(1964)]{mcgee64} McGee, R. X. \& Milton, J. A. 1964, AuJPh,
    17, 128

\bibitem[Mouschovias et al.(1974)]{mouschovias74} Mouschovias, T. Ch., Shu, F. H.,
    Woodward, P. R. 1974, A\&A, 33, 73

\bibitem[Nagasawa(1987)]{nagasawa87} Nagasawa, M. 1987, Prog. Theor. Phys., 77, 635

\bibitem[Nakamura et al.(1993)]{nakamura93} Nakamura, F., Hanawa, T., \& Nakano, T.
    1993, PASJ, 45, 551

\bibitem[Peterken et al.(2019)]{peterken19} Peterken, T.G., Merrifield, M.R.,
    Arag\'on-Salamanca, A., Drory, N., Krawczyk, C.M., Masters, K.L., Weijmans,
    A.-M., Westfall, K.B. 2019, NatAs, 3, 178

\bibitem[Ragan et al.(2014)]{ragan14} Ragan S. E., Henning T., Tackenberg J., et al.
    2014, A\&A, 568, A73

\bibitem[Renaud et al.(2013)]{renaud13} Renaud, F., Bournaud, F., Emsellem, E., et
    al.  2013, MNRAS, 436, 1836

\bibitem[Renaud et al.(2014)]{renaud14} Renaud, F., Bournaud, F., Emsellem, E.,
    Elmegreen, B., \&  Teyssier, R. 2014, ASPC, 480, 247

\bibitem[Roberts(1969)]{roberts69} Roberts, W.W. 1969, ApJ, 158, 123

\bibitem[Rosanno(1978)]{rossano78} Rossano, G. S. 1978, AJ, 83, 234

\bibitem[Rozas et al.(1996)]{rozas96} Rozas, M., Beckman, J. E., Knapen, J. H.
    1996, A\&A, 307, 735

\bibitem[Schneider \& Elmegreen(1979)]{schneider79} Schneider, S., \& Elmegreen,
    B.G. 1979, ApJS, 41, 87

\bibitem[Scoville et al.(1983)]{scoville83} Scoville, N. Z., Sanders, D. B.,
    Clemens, D.P. 1986, ApJ, 310, L77

\bibitem[Seo \& Youdin (2016)]{seo16} Seo, Y.M. \& Youdin, A.N. 2016, MNRAS, 461,
    1088

\bibitem[Shabani et al.(2018)]{shabani18} Shabani, F., Grebel, E.K., Pasquali, A.
    et
    al. 2018, MNRAS, 478, 3590

\bibitem[Shetty \& Ostriker(2006)]{shetty06} Shetty R., \& Ostriker E. C., 2006,
    ApJ, 647, 997

\bibitem[Shu(2016)]{shu16} Shu F. H., 2016, ARA\&A, 54, 667

\bibitem[Stern et al.(2005)]{stern05} Stern, D., Eisenhardt, P., Gorjian, V. et al.
    2005, ApJ, 631, 163

\bibitem[Stod\'o{\l}kiewicz(1963)]{stod63} Stod\'o{\l}kiewicz, J.S. 1963, Acta.
    Astron. 13, 30

\bibitem[Stutz et al.(2013)]{stutz13} Stutz, A., Tobin, J., Stanke, T., Megeath, T.,
    Fischer, W. et al. 2013, ApJ, 767, 36

\bibitem[Thilker et al.(2007)]{thilker07} Thilker, D.A., Bianchi, L., Meurer, G. et
    al. 2007, ApJS, 173, 538

\bibitem[Tomisaka(1987)]{tomisaka87} Tomisaka, K. 1987, PASJ, 39, 109

\bibitem[Tomisaka(1995)]{tomisaka95} Tomisaka, K. 1995, ApJ, 438, 226

\bibitem[Toomre(1981)]{toomre81} Toomre, A. 1981, in: The structure and evolution of
    normal galaxies, eds. S. M. Fall \& D. Lynden-Bell, Cambridge University Press,
    111

\bibitem[Yu \& Ho(2019)]{yu19} Yu, S.-Y., \& Ho, L.C. 2019, ApJ, 871, 194

\bibitem[Wada \& Koda(2004)]{wada04} Wada K. \& Koda J., 2004, MNRAS, 349, 270


\bibitem[Whitney et al.(2003)]{whitney03} Whitney, B. A., Wood, K., Bjorkman, J. E.,
    \& Cohen, M. 2003, ApJ, 598, 1079

\bibitem[Wilson et al.(2019)]{wilson19} Wilson, C.D., Elmegreen, B.G., Bemis, A., \&
    Brunetti, N. 2019, ApJ, 882, 5

\bibitem[Winston et al.(2007)]{winston07} Winston, E., Megeath, S. T., Wolk, S.
    J., et al. 2007, ApJ, 669, 493

\bibitem[Xu et al.(2001)]{xu01} Xu, C., Lonsdale, C.J., Shupe, D.L., O'Linger, J.,
    \& Masci, F. 2001, ApJ, 562, 179

\end{thebibliography}
\end{document}